\documentclass[12pt]{article}

\usepackage{amscd, amsmath}
\usepackage{amsthm, amssymb}
\usepackage{array}

\newcommand{\CC}{\mathbb{C}}

\newcommand{\OO}{\mathcal O}
\newcommand{\he}{\hat{e}}

\newcommand{\db}{\bar{\partial}}
\newcommand{\dd}{{\rm d}}

\newcommand{\g}{\mathfrak g}

\newcommand{\ver}{\ker \dd \pi_E}
\newcommand{\gp}{{\mathfrak g}_P}

\newcommand{\vp}{\omega}

\newcommand{\psip}{\psi_{+}}
\newcommand{\psim}{\psi_{-}}
\newcommand{\psibp}{\bar{\psi}_+}
\newcommand{\psibm}{\bar{\psi}_-}

\newcommand{\lam}{\lambda}

\newcommand{\lamp}{\lambda_{+}}
\newcommand{\lamm}{\lambda_{-}}
\newcommand{\lambp}{\bar{\lambda}_+}
\newcommand{\lambm}{\bar{\lambda}_-}
\newcommand{\la}{\xi}

\newcommand{\zb}{\bar{z}}

\newcommand{\epsilonbp}{\bar{\epsilon}_+}
\newcommand{\epsilonbm}{\bar{\epsilon}_-}
\newcommand{\epsilonp}{\epsilon_+}
\newcommand{\epsilonm}{\epsilon_-}
\newcommand{\phinablaA}{\phi^\ast \nabla^A}
\newcommand{\si}{\sigma}
\newcommand{\sib}{\bar{\sigma}}
\newcommand{\ov}{\overline}
\newcommand{\sq}{\sqrt{2}}

\begin{document}

\begin{titlepage}
\title{
\vskip -70pt
\begin{flushright}
{\normalsize \ ITFA-2007-31}\\
\end{flushright}
\vskip 45pt
{\bf Twisting gauged non-linear sigma-models}
}
\vspace{3cm}

\author{{J. M. Baptista} \thanks{ e-mail address:
    jbaptist@science.uva.nl}  \\
{\normalsize {\sl Institute for Theoretical Physics} \thanks{ address: Valckenierstraat 65, 1018 XE Amsterdam, The Netherlands}
} \\
{\normalsize {\sl University of Amsterdam}} 
}

\date{July 2007}

\maketitle

\thispagestyle{empty}
\vspace{2cm}
\vskip 20pt
{\centerline{{\large \bf{Abstract}}}}
\vspace{.35cm}
We consider gauged sigma-models from a Riemann surface into a K\"ahler and hamiltonian $G$-manifold $X$. 
The supersymmetric ${\mathcal N}=2$ theory can always be twisted to produce a gauged A-model. This model
localizes to the moduli space of solutions of the vortex equations and computes the Hamiltonian Gromov-Witten 
invariants. When the target is equivariantly Calabi-Yau, i.e. when its first $G$-equivariant Chern class 
vanishes, the supersymmetric theory can also be twisted into a gauged B-model. This model localizes to the
K\"ahler quotient $X/\!/ G$.  
\end{titlepage}

\tableofcontents

\section{Introduction}

  Topological field theories are a major tool to explore complex and symplectic geometry. The earliest and most 
well-known examples of their usefulness, dating from almost twenty years ago, were  applications of topological 
sigma-models and mirror symmetry to predict Gromov-Witten invariants of Calabi-Yau manifolds. Since then the subject
has developed in several directions: in depth and rigour, with the invention of new computational techniques for GW-invariants
and mathematical frameworks for mirror symmetry; and in breadth and diversity,  with the discovery of new invariants
and dualities through the use of topological strings and other TFT's.

One such recent development, so far still relatively unexplored, was the definition in the mathematical literature of the 
Hamiltonian Gromov-Witten invariants \cite{C-G-M-S}. These invariants study K\"ahler manifolds equipped with hamiltonian actions of 
compact Lie groups. To define them one uses the moduli space of solutions of the vortex equations. In the special case of 
a trivial group the vortex equations reduce to the equations for holomorphic curves, and hence in this instance the HGW-invariants
reduce to the GW-invariants. Thus these new invariants were introduced as a generalization of the GW-invariants designed to
study hamiltonian actions on symplectic manifolds; moreover, they also bear natural relations with the original 
GW-invariants \cite{Ga-Sa} (see also below). 

From a physics point of view, the HGW-invariants clearly must come from supersymmetric and topological 
gauged non-linear sigma-models. As far as the 
author is aware, however, there is not much literature on this subject. A first motivation for this paper is thus to provide  a framework
to study the HGW-invariants within topological field theory. This is done by considering the ${\mathcal N}=2$ gauged
non-linear sigma-model and, through the usual procedure, twist it to obtain topological gauged A and B models. Since in the
non-gauged case the physical approach, as mentioned above, has been very successful in giving predictions and insights into 
Gromov-Witten theory, we are curious to know how much of this extends to the gauged theories.

A second motivation for this study comes from the fact that, even if one is not interested in the HGW-invariants for
themselves, the gauged sigma-model with target $X$ can be used as a tool to investigate the non-gauged model with 
target $X/\!/ G$. This fact was first recognized in the celebrated paper \cite{W4}, where the gauged linear sigma-model
with target $X=\CC^n$ and group $G = U(1)$ was used to study non-gauged sigma-models into weighted projective spaces
and their Calabi-Yau hypersurfaces. This approach shed new light on the Calabi-Yau/Landau-Ginzburg correspondence and,
at the same time, proved useful as a tool to compute the GW-invariants of toric Calabi-Yau's (e.g. \cite{M-P, Hori-al}). Another
application of gauged linear sigma-models was given in \cite{W5}, where this time the target $X = \CC^{kn}$ and group
$G= U(k)$ were used to study the quantum cohomology  of Grassmannians. More recently, in \cite{Hori-To}, the phase structure and
dynamics of these non-abelian linear models have been further analysed. 
Thus a natural question in the subject, and our second motivation,
is to ask how much of this can be extended to non-linear targets $X$, other quotients $X/\!/G$ and other Calabi-Yau's.
In the mathematical literature these matters have received some investigation in \cite{Ga-Sa}, but to the author's knowledge
they have not been addressed on the physics side.

Our purpose in this paper is to give an impulse to these investigations by describing in detail the supersymmetric ${\mathcal N}=2$
gauged non-linear sigma-model, the gauged A and B models, their observables and localization moduli spaces.

$\ $

We now give a rather detailed description of the contents of the paper. We deal with gauged sigma-models, in other words 
theories that couple matter 
and gauge fields. Matter fields are represented by maps $\phi: \Sigma \rightarrow X$ from a Riemann surface into a K\"ahler 
target. Gauge fields are represented by a $G$-connection $A$ over the Riemann surface. In order to couple these two fields 
one also assumes that the gauge group $G$ acts on the target $X$ in a holomorphic and hamiltonian way. The most important 
part of the action of these models is then
\begin{equation}
I(A, \phi) = \int_\Sigma  |F_A|^2 + |\dd^A \phi|^2 + |\mu \circ \phi|^2 + \cdots  \ ,
\label{1.1}
\end{equation}
where $F_A$ is the curvature of $A$, $\dd^A \phi$ is a covariant derivative and $\mu : X \rightarrow \text{Lie }G$ 
is the moment map of the $G$-action. This action reduces to the classical action of sigma-models if we take $G$ to 
be trivial. Now, the usual sigma-models have ${\mathcal N}=2$ supersymmetric extensions for K\"ahler target $X$. It is then 
a fact that, when $X$ has a group $G$ of hamiltoniam isometries, the ${\mathcal N}=2$ theory can be gauged while 
preserving the supersymmetry, i.e. (\ref{1.1}) has a ${\mathcal N}=2$ supersymmetric extension. Similarly,  
the topological theories that will be described here --- the gauged A and B models --- are both extensions of (\ref{1.1})
 obtained by considering extra fields and adding more terms to the action. In fact there are basically two standard ways of 
constructing this kind of topological theories: one is by twisting the supersymmetric theory mentioned before; the second is 
through the use of the Mathai-Quillen formalism. The latter has a more geometric flavour and was already applied in 
\cite{Bap} to the gauged A-model. Twisting 
the supersymmetric theory, on the other hand, not only is more familiar a method to the physicists, but also has the
advantage that, in the non-gauged case, produces two distinct and equally important topological theories: the A and
B models. This does not happen with the Mathai-Quillen formalism, which only yields the A-theory. Since in this
paper our main aim is to extend both models to the gauged case, we will proceed through the twist. We wish to stress
that all these twisting constructions are very standard in the non-gauged case, and thus, since things are quite 
similar here, we present most of the results without detailed calculations. We took some trouble, nevertheless, in 
trying to present consistent and detailed formulas. 

$\ $

In section 2 we spell out the fields, action and supersymmetry transformations of the ${\mathcal N}=2$ gauged 
non-linear sigma-model in two dimensions. These are obtained by dimensional reduction of the ${\mathcal N}=1$
gauged non-linear sigma-model in four dimensions \cite{D-F}. This supersymmetric ${\mathcal N}=2$ model, like its non-gauged
counterpart, possesses two classical $U(1)$-symmetries. Standard index theorems are then applied to show that
one of them, the vectorial R-symmetry, is always non-anomalous, whereas the other one, the axial R-symmetry, is in 
general anomalous. Sometimes, however, the axial anomaly also vanishes, and a sufficient condition for this to 
happen is that $c_1^G (TX)$, the first $G$-equivariant Chern class of $X$, vanishes. Targets $X$ with this 
property are called equivariant Calabi-Yau's; they may also be characterized by the fact that they possess a 
$G$-invariant and nowhere-vanishing $(n,0)$-form, where $n$ is the complex dimension of $X$. Now, since the twists
of the supersymmetric theory are performed along the non-anomalous R-symmetries, manifolds with $c_1^G (TX) =0$
are very special, for they support two distinct twisted theories, the gauged A and B models. A general K\"ahler
target, on the other hand, only supports the gauged A-model. 
 A pleasant property of equivariant Calabi-Yau's, we find, is that their K\"ahler quotient $X/\!/G$ is also
Calabi-Yau. Three simple examples of equivariant Calabi-Yau's are presented at the end of section 2.2. The first
is complex vector spaces with special unitary representations of $G$. The second is when $X$ is the total space of a 
sum of line-bundles 
over a complex base, $X = \oplus_k L_k \rightarrow M$, with the circle $U(1)$ acting on each line-bundle with charge
$q_k$, and with the two algebraic conditions $\sum_k q_k = 0$ and $c_1 (TM) + \sum_k c_1 (L_k) = 0$ satisfied. 
The third example is hyperk\"ahler manifolds with compatible $G$-actions, and we give a short list of famous spaces of this sort 
at the end of 2.2.

While all of these are well known examples of Calabi-Yau's, it is not obvious to the author whether all of them, 
or their quotients, 
can be studied within the framework of the gauged linear models, i.e. as hypersurfaces or complete intersections in toric
varieties or Grassmanians. If this is not the case, then there may be some scope for these models as tools to investigate
Calabi-Yau's; the hope is that, just as in 
the linear case, some aspects of the theories may be easier to study in their gauged (or unquotiented) version than in 
the ungauged version on the quotient space. At least aspects related to the phase structure of the theory and, more ambitiously, 
to mirror symmetry, seem to fit well with gauged theories \cite{W4, Hori-al}. Another point of view would be to be less concerned
about the quotients and just decide to study hamiltonian actions on symplectic manifolds, in which case the framework of
non-linear gauged sigma-models and HGW-invariants is the appropriate one.

$\ $

In section 3 we turn to the topological theories, starting with the gauged A-model. The fields, action and $Q_A$-operator are written down in 
the explicit formulas derived from the supersymmetric theory. These formulas had already been obtained in \cite{Bap}
through the Mathai-Quillen formalism. (The material in this section, in fact, is almost entirely contained either
in \cite{Bap} or in \cite{W2}, and so the section can be regarded as a review, or at most a check that the supersymmetric twist
agrees with the Mathai-Quillen result.) Concerning the observables of the theory, recall that in the non-gauged A-model
they are constructed using de Rham cohomology classes of the target $X$; in the gauged A-model, not surprisingly, they
are constructed via the $G$-equivariant cohomology of $X$. In the non-gauged theory, moreover, the path-integrals 
that compute the expectation values of the observables get localized to integrals over the finite-dimensional space
of holomorphic curves; in the gauged A-model, on the other hand, the localization is to the moduli space of solutions of 
the general vortex equations. These expectation values are then closely related to the Hamiltonian Gromov-Witten 
invariants of $X$ \cite{Bap, C-G-M-S}, a type of invariants that studies vortex moduli spaces and generalizes 
(at least part of) the usual Gromov-Witten theory.

$\ $

Finally in section 4 we look at the B-twist of the gauged supersymmetric theory. Since this topological theory is
not accessible through the Mathai-Quillen formalism, it was not considered in \cite{Bap}, and also seems not to have been
much studied anywhere else. In section 4 we spell out the fields, action and $Q_B$-transformations of this theory. In this section
we include in the theory a non-zero superpotential
$W$ (in itself just a $G$-invariant and holomorphic function on $X$), and thus obtain a gauged Landau-Ginzburg model.
As in the non-gauged case this is possible because a non-zero $W$ does not spoil the axial symmetry (used, recall, 
to define the B-twist), whereas it usually spoils the vector symmetry.
Regarding localization, it is argued in section 4.2 that the path-integrals of the B-theory localize to a set of 
field configurations that is smaller than the set of $Q_B$-fixed points. This is related to the usual decomposition 
$Q_B = \ov{Q}_+ + \ov{Q}_-$ and to the fact that the B-action is simultaneously $\ov{Q}_+$- and $\ov{Q}_-$-exact, up to
topological terms. It is then shown that in favourable cases, including whenever $\Sigma$ has genus zero, this 
smaller set can be identified with the K\"ahler quotient $X/\!/G$, which, as said before, is Calabi-Yau. For 
a non-zero superpotential $W$ the localization set is furthermore restricted to the critical set of $W$ in $X/\!/G$.

$\ $

  Generally speaking the work in this paper extends in the natural way some classical aspects of the 
non-gauged and the gauged linear sigma-models. Not addressed here are the important quantum aspects of the theory, especially 
the RG-flow, the $\beta$-function and the singularities in the Fayet-Iliopoulos parameter space (here equal to 
the center of $\g$). For instance, naively extending \cite{W4}, one would expect that $\beta =0$ at one loop for equivariant
Calabi-Yau's and that the quantum singularities will appear for values of the FI-parameter such that the set $\mu^{-1}(0)$ 
contains points with non-trivial $G$-stabilizer. As in \cite{W4}, the analysis of these problems is essential to investigate
the existence of Calabi-Yau/Landau-Ginzburg correspondences and mirror dualities in these gauged models. Another possible
direction is to study the coupling of these gauged sigma-models to gravity, or, maybe better, to the complex structure of the 
worldsheet.

\section{The gauged ${\mathcal N}=2$ supersymmetric theory}

\subsection{Fields, lagrangian and supersymmetries}

In two dimensions, globally supersymmetric theories are defined only on flat spacetimes, so in this section we take $\Sigma$
to be either the complex plane, the cylinder or the torus.
The two main fields of the gauged sigma-model are a connection $A$ on a principal $G$-bundle $P \rightarrow \Sigma$
and a section $\phi: \Sigma \rightarrow E$ of the associated bundle $E := P\times_G X$. Observe that locally $E$ looks
like the product $\Sigma \times X$, and so locally $\phi$ looks like a map $\Sigma \rightarrow X$. This is globally 
true when $P$ is the trivial $G$-bundle. 
Besides the scalar section $\phi$ and the connection $A$, the other fields of the supersymmetric theory are:
\begin{align}
\si &\in \Omega_+^0 (\Sigma ; \gp^\CC )   &     F &\in \Omega_+^0 (\Sigma ; \phi^\ast \ver )    \label{2.1}   \\
\psi_{\pm}  &\in  \Omega_-^0 (\Sigma ;  S_{\pm} \otimes \phi^\ast \ver )  &      D  &\in \Omega_+^0 (\Sigma ; \gp ) \nonumber  \\
\lam_{\pm}  &\in  \Omega_-^0 (\Sigma ;  S_{\pm} \otimes \gp^\CC )   \nonumber
\end{align}
Here, as in the rest of the paper, the notation $\Omega^p_{\pm} (\Sigma ; V)$ represents the space of $p$-forms
on $\Sigma$ with values on the bundle $V \rightarrow \Sigma$; the signs in subscript distinguish bosonic fields (+)
from fermionic ones (--). The bundles that appear in (\ref{2.1}) are: the adjoint bundle $\gp := P\times_{\text{Ad}} \g$ ---  where 
$\g$ denotes the Lie algebra of $G$ ---  and its complexification $\gp^\CC$; the spinor bundles of the Riemann surface
$S_{\pm} = K^{\pm 1/2}$, with $K = \Lambda^{1,0}\Sigma$ being the canonical bundle of $\Sigma$; the bundle 
$\ver \rightarrow E$, which locally looks like $\Sigma \times TX \rightarrow \Sigma \times X$, and is just the
sub-bundle of $TE \rightarrow E$ defined as  the kernel of the derivative of the projection $\pi_E : E \rightarrow 
\Sigma$; and finally $\phi^\ast (\ver) \rightarrow \Sigma$, the pull-back of $\ver$ by the 
section $\phi$. Thus in the end we have one adjoint scalar field $\si$, four fermionic fields $\psi_\pm$ and
$\lambda_\pm$, and two scalar auxiliary fields $F$ and $D$.

Using all these fields one can define the Lagrangian of the euclidean supersymmetric theory as
\begin{equation}
L_{\text{SUSY}} =  L_{\text{matter}} +  L_{\text{gauge}} + L_W   +   L_{\theta, B}   \ ,
\label{2.2}
\end{equation}
where the various components are as follows. The matter part, which upon putting $A= 0$ reduces to the lagrangian
of the non-gauged sigma-model, is
\begin{equation*}
\begin{split}
L_{\text{matter}} =& 
|\dd^A \phi|^2 \:+\: |\si^a \he_a |^2 \:+\: |\sib^a \he_a|^2 \:+\: 
2\,i\, h_{j \bar{k}}\, \ov{\psip^k} \, (\phinablaA)_z \psip^j  
\:-\: 2\,i\, h_{j\bar{k}} \,\ov{\psim^k} \, (\phinablaA)_{\zb} \psim^j  \\
&\:-\: \sqrt{2}\, i \, h_{j\bar{k}} \, (\nabla_l \, \he_a^j)\, (\si^a \ov{\psim^k} \psip^l  \:+\: 
\sib^a \ov{\psip^k} \psim^l )  \:+\:  R_{i\bar{j}k\bar{l}} \, \psip^i \psim^k \ov{\psim^j}\, \ov{\psip^l}   \\
& \:-\:  \sqrt{2} \, h_{j\bar{k}} \,(\ov{\lamp^a}\, \he_a^j \, \ov{\psim^k} \:-\: \ov{\lamm^a}\, \he_a^j\, \ov{\psip^k} 
\:-\: \lamp^a\, \ov{\he_a^k}\, \psim^j \:+\: \lamm^a\, \ov{\he_a^k}\,  \psip^j)  \nonumber   \\
& \:-\:  h_{j\bar{k}} (F^j \:-\: \Gamma^j_{il}\, \psip^i \psim^l) (\ov{F^k} \:-\: \ov{\Gamma^k_{mn}}\, \,
\ov{\psim^m}\, \, \ov{\psip^n})   \ .
\end{split}
\end{equation*}
Here $\{ e_a \}$ denotes a basis of the Lie algebra $\g$ and $\he_a$ the vector field on $X$ associated to 
$e_a$ by the left $G$-action. The lagrangian $L_{\text{gauge}}$, which upon putting $X = \text{point}$ reduces to
the pure Yang-Mills lagrangian, is
\begin{equation*}
\begin{split}
L_{\text{gauge}}  =&
\frac{1}{e^2}  \Bigl\{  \frac{1}{2}\, | F_A |^2 \: + \:  | \dd^A \sigma |^2 \: + \: \frac{1}{2} \, 
| [\si , \sib] |^2 \: - \: \frac{1}{2} \, |D|^2 \: + \: 2\, e^2 \, \phi^\ast \mu_a \, D^a \\
\:&+ \:  2\,i \,(\lambp)_a \nabla_{z}^A \lamp^a \: - \: 2\,i\, (\lambm)_a \nabla^A_{\zb} \lamm^a \; 
- \;\sqrt{2} \,i\, \lambm^a [\si , \lamp]_a  \: - \: \sqrt{2}\, i\, \lambp^a [\sib , \lamm]_a  \Bigr\}   \ ,
\end{split}
\end{equation*}
where $\mu : X \rightarrow \g^\ast$ is a moment map of the $G$-action on $X$ (for the standard definition of 
$\mu$ see appendix A). The superpotential term is
\begin{equation*}
L_W  =    \frac{1}{2} \,  F^k  (\partial_k W) +  \frac{1}{2} \, \psim^j \psip^k  (\partial_j \partial_k W) + 
\frac{1}{2} \, \ov{F^k}  (\partial_{\bar{k}} \ov{W}) +  \frac{1}{2}\, \ov{\psip^k}\, \, \ov{\psim^j}  \, 
(\partial_{\bar{j}} \partial_{\bar{k}}\ov{W})  \ ,
\end{equation*}
where $W$, the superpotential, is a fixed, non-dynamical, $G$-invariant and holomorphic function on $X$.
Notice that if $X$ is compact only $L_W =0$ is possible. Finally the theta and B-field terms are
\begin{equation*}
L_{\theta, B} =     i \, \phi^\ast B \;  - \; \frac{i}{2\pi} \, (\theta ,  F_A )  \ ,
\end{equation*}
where $B$ is an arbitrary, but fixed, $G$-invariant and closed 2-form on $X$\footnote{In fact the supersymmetric ${\mathcal N} =(2,2)$ 
theory admits a more general $H$-flux term, instead of the B-field term presented here.
This is related to the fact that it also admits more general targets $X$, namely (twisted) generalized K\"ahler manifolds, 
instead of just the K\"ahler targets to which we have restricted ourselves here. For these matters see \cite{K-T} and the 
references therein.};   
$\theta$ is a constant\footnote{Recall that the moment map $\mu$ is also 
defined only up to a constant in  $[\g, \g]^0$, so that both these constants can be combined into an element of the complexified 
space $[\g, \g]^0_\CC$. This complex constant, as usual, is the important parameter of the quantum theory. Note, moreover, that the inner product 
$\kappa$ allows the identification of $[\g, \g]^0$ with the centre of $\g$.}  
in $[\g ,\g]^0$, the subspace of $\g^\ast$ that annihilates commutators; and $(\cdot , \cdot )$ is the natural pairing 
$\g^\ast \times \g \rightarrow {\mathbb R}$.

$\ $

The supersymmetric lagrangian  (\ref{2.2}) is a ${\mathcal N} =(2,2)$ lagrangian, and so is invariant (up to total derivatives)
under four independent fermionic symmetries, whose parameters are denoted $\epsilon_\pm$ and $\bar{\epsilon}_\pm$.
The general supersymmetry transformations are, for the matter fields,  
\begin{align}
\delta \phi^k &=  \sqrt{2} (\epsilon_{+} \psim^k - \epsilon_{-} \psip^k )    \label{2.3}  \\
\delta \ov{\phi^k} &= - \sqrt{2} (\epsilonbp \ov{\psim^k} - \epsilonbm \ov{\psip^k} )  \nonumber  \\
\delta \psip^k &= 2 \sqrt{2} i \epsilonbm (\dd^A_{\zb} \phi^k ) + \sqrt{2} \epsilon_+ F^k +  2 i \epsilonbp \sib^a \he_a^k   \nonumber \\ 
\delta \ov{\psip^k} &= - 2 \sqrt{2} i \epsilon_- (\dd^A_{\zb} \ov{\phi^k} ) + \sqrt{2} \epsilonbp \ov{F^k } 
- 2 i \epsilon_+ \si^a \ov{\he_{a}^{k}}    \nonumber \\
\delta \psim^k &= 2 \sqrt{2} i \epsilonbp (\dd^A_{z} \phi^k ) + \sqrt{2} \epsilon_- F^k -  2 i \epsilonbm \si^a \he_a^k   \nonumber \\ 
\delta \ov{\psim^k} &= - 2 \sqrt{2} i \epsilon_+ (\dd^A_{z} \ov{\phi^k} ) + \sqrt{2} \epsilonbm \ov{F^k } 
+ 2 i \epsilon_- \sib^a \ov{\he_{a}^{k}}    \nonumber \\
\delta F^k  &=  2 \sqrt{2} i \epsilonbp (\partial_z \psip^k + A^a_z (\partial_j \he_a^k) \psip^j )   
- 2 \sqrt{2} i \epsilonbm (\partial_{\zb} \psim^k + A^a_{\zb} (\partial_j \he_a^k) \psim^j )   \nonumber \\
&\ \ + 2 \epsilonbm \ov{\lam^a_+} \, \he_a^k -   2 \epsilonbp \ov{\lam^a_-}\, \he_a^k - 2 i \epsilonbp \sib^a (\partial_j \he_a^k) \psim^j 
- 2 i \epsilonbm \si^a (\partial_j \he_a^k) \psip^j   \nonumber \\
\delta \ov{F^k}  &=  2 \sqrt{2} i \epsilonp (\partial_z \ov{\psip^k} + A^a_z (\ov{\partial_j \he_a^k}) \ov{\psip^j} )   
- 2 \sqrt{2} i \epsilonm (\partial_{\zb} \ov{\psim^k} + A^a_{\zb} (\ov{\partial_j \he_a^k}) \ov{\psim^j} )   \nonumber \\
&\ \ - 2 \epsilonm \lam^a_+ \, \ov{\he_a^k} +  2 \epsilonp \lam^a_- \, \ov{\he_a^k} - 2 i \epsilonp \si^a (\ov{\partial_j \he_a^k}) \ov{\psim^j} 
- 2 i \epsilonm \sib^a (\ov{\partial_j \he_a^k}) \ov{\psip^j}  \ . \nonumber 
\end{align}
The gauge fields, at the same time, transform as
\begin{align}
\delta A_z^a  &=  - i \epsilonm \lambm^a -  i \epsilonbm \lamm^a     \label{2.4}    \\
\delta A_{\zb}^a &=  i \epsilonp \lambp^a + i \epsilonbp \lamp^a    \nonumber \\
\delta \si^a &=  -  \sqrt{2} i \epsilonbp \lamm^a - \sqrt{2} i \epsilonm \lambp^a    \nonumber \\
\delta \sib^a &=  -  \sqrt{2} i \epsilonp \lambm^a - \sqrt{2} i \epsilonbm \lamp^a    \nonumber \\
\delta \lamp^a  &=  2\sqrt{2} \epsilonm (\nabla_{\zb}^A \sib^a ) + \epsilonp (i (F_A)_{12}^a  + [\si , \sib]^a + i D^a  )   \nonumber \\
\delta \lambp^a  &=  2\sqrt{2} \epsilonbm (\nabla_{\zb}^A \si^a ) + \epsilonbp (i (F_A)_{12}^a  - [\si , \sib]^a - i D^a  )   \nonumber \\
\delta \lamm^a  &=  - 2\sqrt{2} \epsilonp (\nabla_{z}^A \si^a ) + \epsilonm (- i (F_A)_{12}^a  - [\si , \sib]^a + i D^a  )   \nonumber \\
\delta \lambm^a  &=  - 2\sqrt{2} \epsilonbp (\nabla_{z}^A \sib^a ) + \epsilonbm (- i (F_A)_{12}^a  + [\si , \sib]^a - i D^a  )   \nonumber \\
\delta D^a  &=  2 \epsilonbp (\nabla^A_z \lamp^a ) - 2 \epsilonbm ( \nabla^A_{\zb} \lamm^a ) -  2 \epsilonp (\nabla^A_z \lambp^a )  
+ 2 \epsilonm (\nabla^A_{\zb} \lambm^a )    \nonumber \\
& \ \ +  \sqrt{2} \epsilonp [\si , \lambm]^a  +  \sqrt{2} \epsilonm [\sib , \lambp]^a 
-  \sqrt{2} \epsilonbp [\sib , \lamm]^a  -  \sqrt{2} \epsilonbm [\si , \lamp]^a    \ . \nonumber 
\end{align}

In the lagrangian and supersymmetry transformations written above we have made use of the covariant derivatives induced by $A$ on the 
 bundles $E$, $\gp$ and $\phi^\ast \ver$ over $\Sigma$. These covariant derivatives have the local form
\begin{align}
\dd^A \phi^k  &= \dd \phi^k  +  A^a \, \he^k_a   \label{2.13}   \\
\nabla^A \sigma^a  &=  \dd\sigma^a  +  [A , \sigma]^a     \nonumber \\
(\phi^\ast \nabla^A) \psi^k &=  \dd \psi^k  +  A^a \psi^j   \nabla_j \he^k_a  + \Gamma^k_{jl} (\dd\phi^j) \psi^l  \ , \nonumber
\end{align}
where $\phi$ is locally regarded as a map $\Sigma \rightarrow X$, $\sigma$ as a map $\Sigma \rightarrow \g$, $\psi$ as a 
(fermionic) map $\Sigma \rightarrow \phi^\ast TX$ and $A$ as a local 1-form on $\Sigma$.

\subsection{R-symmetries, anomalies and equivariant Calabi-Yau's}

\subsubsection*{The vector and axial symmetries}

The gauged supersymmetric lagrangian (\ref{2.2}) has, as usual, more symmetries besides the galilean, gauge and 
supersymmetry invariances. These are the two $U(1)$-symmetries called vector and axial R-symmetries.
The vector symmetry is
\begin{align}
\psi_{\pm}  &\longrightarrow e^{-i\alpha}  \psi_{\pm}  &    F &\longrightarrow e^{-2i\alpha} F   \label{2.5}   \\
\lam_{\pm}  &\longrightarrow e^{i\alpha}  \lam_{\pm} \ ,  &  & \nonumber 
\end{align}
with the conjugate fields transforming in the conjugate representation and all other fields remaining invariant.
The axial symmetry is 
\begin{align}
(\psip , \lamp)  &\longrightarrow e^{-i\alpha}  (\psip , \lamp)  &    \si &\longrightarrow e^{2i\alpha} \si  \label{2.6} \\
(\psim , \lamm )  &\longrightarrow e^{i\alpha}  (\psim , \lamm) \ ,  &  & \nonumber 
\end{align}
with, again, the conjugate fields transforming in the conjugate representation and all other fields remaining invariant.

A priori these R-symmetries are only symmetries of the classical theory. To decide whether they are also symmetries of 
the quantum theory, i.e. whether they preserve the measure of the path-integral, one should, as usual, look at the 
kinetic terms of the fermions and analyse their zero-modes. In our case the relevant kinetic terms of the supersymmetric 
lagrangian are
\begin{equation*} 
2\,i\, h_{j \bar{k}}\, \ov{\psip^k} \, (\phinablaA)_z \psip^j  
\:-\: 2\,i\, h_{j\bar{k}} \,\ov{\psim^k} \, (\phinablaA)_{\zb} \psim^j
\: + \:  2\,i \,(\lambp)_a \nabla_{z}^A \lamp^a \: - \: 2\,i\, (\lambm)_a \nabla^A_{\zb} \lamm^a  \ ,
\end{equation*} 
and thus,  for example,
\begin{equation*}
\# \{ \psip\ {\rm zero\ modes} \}\: =\:  {\rm dim\ ker}(\phi^\ast \nabla^A)_z \ .
\end{equation*}
Calculating on the compact torus, Stokes' theorem also allows one to write
\begin{equation*}
\int_{T^2} 2\,i\, h_{j \bar{k}}\, \ov{\psip^k} \, (\phinablaA)_z \psip^j \: = \: 
\int_{T^2} 2\,i\, h_{j \bar{k}}\, \psip^j \, \ov{(\phinablaA)_{\zb} \psip^k } \ ,
\end{equation*}
so that $(\phinablaA)_{\zb}$ is the adjoint operator of $(\phinablaA)_z$ and
\begin{equation*}
\# \{ \ov{\psip} {\rm\ zero\ modes} \}\: =\:  {\rm dim\ ker}(\phi^\ast \nabla^A)_{\zb}\: =\: {\rm dim\ coker}(\phi^\ast \nabla^A)_z \ .
\end{equation*}
Similar calculations determine the number of zero modes of the other fermionic fields. Now,  the standard heuristic analysis 
of the path-integral measure says that if a fermion field $\chi$ is acted by a $U(1)$-symmetry with charge $q(\chi)$, then 
the functional measure ${\mathcal D} \chi$ transforms under this symmetry with a charge $-q(\chi)$ times the number of 
$\chi$ zero modes. This means in our examples that
\begin{equation*}
{\mathcal D}\psi_{\pm} {\mathcal D}\ov{\psi_\pm} {\mathcal D}\lambda_\pm {\mathcal D}\ov{\lambda_\pm}
\: \longrightarrow \:  e^{-i{\mathcal A}\alpha} \, 
{\mathcal D}\psi_{\pm} {\mathcal D}\ov{\psi_\pm} {\mathcal D}\lambda_\pm {\mathcal D}\ov{\lambda_\pm} \ ,
\end{equation*}
where the anomaly ${\mathcal A}$ is
\begin{equation*}
{\mathcal A} \: =\:  [q(\psim) - q(\psip)] ( {\rm index }\ \phi^\ast \nabla^A_{\zb}) + 
[q(\lamm) - q(\lamp)] ( {\rm index }\  \nabla^A_{\zb})  \ .
\end{equation*} 
Notice that this quantity automatically vanishes for the vector symmetry (\ref{2.5}), as expected, and so also in the gauged 
model  this symmetry is non-anomalous. As for the axial symmmetry, its anomaly depends on the index
of the Cauchy-Riemann operators 
\begin{align}
(\nabla^A)^{0,1} &:  \Omega^0 (\Sigma ; \gp)  \longrightarrow \Omega^{0,1} (\Sigma ; \gp)  \label{2.7}  \\
(\phi^{\ast}\nabla^A)^{0,1} &:  \Omega^0 (\Sigma ; \phi^\ast \ver)  \longrightarrow \Omega^{0,1} (\Sigma ; \phi^\ast \ver) \ . \nonumber 
\end{align}
This index is easily obtained from the Hirzebruch-Riemann-Roch theorem, and the result for a general compact $\Sigma$ is
\begin{align}
\rm{index} (\nabla^A)^{0,1} &=  c_1 (\gp \rightarrow \Sigma) +  ({\rm dim}G)(1-g)   \label{2.8}  \\
\rm{index} (\phi^{\ast}\nabla^A)^{0,1}   &=  c_1 (\phi^\ast \ver \rightarrow \Sigma) +  ({\rm dim}_\CC X)(1-g) \ . \nonumber 
\end{align}
This is the complex index of the operators. For a compact Lie group, however,  
the Chern number $c_1 (\gp)$ always vanishes, and since we are calculating on a torus the final result for the
axial anomaly is
\begin{equation*}
{\mathcal A}({\rm axial}) = 2 \: c_1 (\phi^\ast \ver \rightarrow \Sigma) = 2 \: \langle  c_1^G (TX)\: ,\: \phi_{\ast} (\Sigma) \rangle \ .
\end{equation*} 
The right-hand-side way of representing the Chern number $ c_1 (\phi^\ast \ver )$ was noted in \cite{C-G-M-S} and requires a 
little explanation. The quantity $c_1^G (TX)$ is the first 
$G$-equivariant Chern
class of the tangent bundle $TX$, and thus belongs to the equivariant cohomology space $H^2_G (X)$; the symbol 
$\phi_{\ast} (\Sigma)$ represents here the equivariant homology class in $H_2^G (X)$ obtained by push-forward by
$\phi$ of the fundamental class of $\Sigma$; finally the brackets are just the natural bilinear pairing
$H^2_G (X) \times H_2^G (X) \rightarrow {\mathbb R}$ (for more details on equivariant cohomology see \cite{B-G-V, G-G-K, Gu-S}). 
The merit of this right-hand-side representation is that it
shows manifestly that a sufficient condition for the axial anomaly to vanish for all $\phi$ is that
\begin{equation}
c_1^G (TX) = 0  \ ,
\label{2.9}
\end{equation}
which may be called the equivariant Calabi-Yau condition.

\subsubsection*{On equivariant Calabi-Yau's}

As is well known, in the usual non-equivariant case the vanishing of the first Chern class is equivalent to the triviality 
of the Ricci class, or, in other words, to the triviality of the canonical bundle.  Similar results hold in the equivariant case.
We will now describe how this goes and, at the end of the section, present two simple examples of equivariant Calabi-Yau's.

Recall that the $G$-equivariant complex $\Omega^\bullet_G (X)$ of the manifold $X$ is, in the Cartan model, the set of
$G$-invariant elements in the tensor product $S^\bullet (\g^\ast) \otimes \Omega^\bullet (X)$. Here $S^\bullet (\g^\ast)$ denotes
the symmetric algebra of $\g^\ast$ and $\Omega^\bullet (X)$ the de Rham complex of $X$. The differential operator of this complex
is $\dd_G = 1 \otimes \dd + e^a \otimes \iota_{\he_a}$, and since $(d_G)^2 = 0$ on elements of $\Omega_G^\bullet (X)$, one can
consider the equivariant cohomology $H_G^\bullet (X)$ of the complex (see, again,  \cite{B-G-V, G-G-K, Gu-S} for more details). 
Now,  according to the results of \cite{B-G-V} and \cite{B-T}, the Chern class $c_1^G (TX)$ is represented in the Cartan model by the 
equivariant form 
\begin{equation}
\eta \ =\  \frac{i}{2\pi} {\rm Tr}^{\CC} (R + e^a \otimes \nabla \he_a )   \  \qquad  \in \ \Omega^2_G (X)\ . 
\label{2.10}
\end{equation}
Here $R$ is the curvature form of the Levi-Civita connection, thus an element of $\Omega^2 (X; {\rm End}_\CC TX)$, and 
$\nabla \he_a$ belongs to $\Omega^0 (X; {\rm End}_\CC TX)$. Notice that on a common Riemannian manifold these forms have 
values on the real endomorphism bundle ${\rm End}_{\mathbb R} TX$; however, when $X$ is K\"ahler and $\he_a$ is holomorphic Killing,
one can show that they actually are $J$-linear and have values on the complex anti-hermitian endomorphisms of $TX$. 
As a representative of a characteristic class, the form $\eta$ must necessarily be $\dd_G$-closed, a fact that can also be 
checked directly. 

A second way of writing (\ref{2.10}) follows from the standard fact $i {\rm Tr}^{\CC} R = \rho$, where $\rho$ denotes the Ricci
form of $X$, and the identities
\begin{equation*}
2i\, {\rm Tr}^{\CC}(\nabla \hat{v}) \: =\: 4\, v^a\, h^{j\bar{k}}  \,\partial_j \partial_{\bar{k}}\, \mu_a \:= \: -(\Delta \mu , v) \ , 
\end{equation*}
which are valid on hamiltonian K\"ahler manifolds. One can thus write\footnote{This formula suggests that the natural analog in the
hamiltonian setting of a Ricci-flat K\"ahler metric is a $G$-invariant Ricci-flat K\"ahler metric whose moment map is harmonic.}
\begin{equation*}
\eta \: =\: \frac{1}{2\pi}\, \rho \: -\:  \frac{1}{4\pi} \, e^a \otimes \Delta \mu_a \ .
\end{equation*}
Finally, the Calabi-Yau condition (\ref{2.9}) is equivalent to the $\dd_G$-exactness of $\eta$, or in other words 
to the existence of a $G$-invariant real form $\sigma \in \Omega^1 (X)$ such that
\begin{align*}
\begin{cases} 
\dd \sigma \, =\,  \rho  &   \\
\iota_{\hat{v}}\sigma \,  =\,   -(\Delta \mu , v)/2  \qquad {\text{for all}} \ v \in \g   \ . &
\end{cases}
\end{align*}

Another (related) characterization of the equivariant Calabi-Yau condition comes from considering the canonical line-bundle
$K = \Lambda^{n,0}X \rightarrow X$, where $n$ is the complex dimension of $X$. This bundle inherits from $X$ a natural $G$-action 
that preserves its natural hermitian metric. It follows from the definitions of \cite{B-G-V} or \cite{B-T} that the $G$-equivariant
 curvature form of $K \rightarrow X$ is 
\begin{equation*}
- i \rho + e^a \otimes {\rm Tr}^\CC (\nabla \he_a) \ ,
\end{equation*}
and therefore that $c_1^G (K) =  c_1^G (TX)$. In particular the Calabi-Yau condition is equivalent to $c_1^G (K) = 0$, and by 
the classification of complex $G$-equivariant line-bundles \cite{MiR}, this is the same as demanding the equivariant triviality 
of $K$. In conclusion, $X$ is equivariantly Calabi-Yau if and only if there exists a nowhere-vanishing and $G$-invariant form
$\Omega \in \Omega^{n,0}(X)$. This form, of course, is unique up to multiplication by nowhere-vanishing $G$-invariant complex 
functions.

$\ $

One pleasant feature of equivariant Calabi-Yau's is their relation to K\"ahler quotients, namely that the quotient of an
equivariant Calabi-Yau is Calabi-Yau. To justify this suppose that $X$ is a K\"ahler manifold equipped with a hamiltonian
and holomorphic $G$-action such that $G$ acts freely on $\mu^{-1} (0)$. Then the K\"ahler quotient $X/\!/G$ exists as a smooth
K\"ahler manifold. If in addition $X$ is equivariantly Calabi-Yau, let $\Omega \in \Omega^{n,0}(X)$ be the $G$-invariant and 
nowhere-vanishing form described above. Then it is not difficult to show that the form
\begin{equation*}
\tilde{\Omega} :=   \sqrt{|\text{det } k_{ab}|} \iota_{\he_1} \cdots  \iota_{\he_r} \Omega \ ,       \qquad r=\text{dim }\g \ ,
\end{equation*}
after restriction to $\mu^{-1}(0)$, descends to a nowhere-vanishing $(n-r)$-form on the quotient $\mu^{-1}(0)/G = X/\!/G$. Using 
the definition of the complex structure on $X/\!/G$ induced by $X$ one can, moreover, verify that this is in fact a $(n-r, 0)$-form, 
and so $X/\!/G$ is Calabi-Yau. A straightforward generalization of this argument shows also that if $H$ is a normal subgroup of $G$,
then the quotient $X/\!/H$ is a $G/H$-equivariant Calabi-Yau.

\subsubsection*{Examples of equivariant Calabi-Yau's}

To close this section we will give a few examples of equivariant Calabi-Yau's. For the first one,
let $X$ be a complex vector space equipped with a hermitian product, and let $r$ be a unitary representation of $G$ on $X$. 
Then $\dd r$, the associated representation of the Lie algebra $\g$, has values on the anti-hermitian endomorphisms of $X$. 
Now, by deformation invariance 
\cite[Appendix C]{G-G-K}, two $\dd_G$-closed forms in $\Omega^\bullet_G (X)$ are cohomologous iff they coincide at the origin of 
the vector space $X$. Therefore $[\eta]_G = 0$ iff
\begin{equation*}
\rho\: |_{\rm origin} = {\rm Tr}^\CC (\nabla \hat{v})\: |_{\rm origin} = 0 \qquad {\text{for all}}\ v \in \g \ .
\end{equation*}
But since $X$ has no curvature, we have that $\rho \equiv 0$ and that
\begin{align*}
\hat{v} &=  [\dd r (v)]^j_k w^k \,  \frac{\partial}{\partial w^j}  \\
(\nabla \hat{v})^j_k   &=   [\dd r (v)]^j_k \ ,
\end{align*}
and so $c_1^G (TX) = 0$ if and only if
\begin{equation*}
{\rm Tr}^\CC  [\dd r (v)] = [\dd r (v)]^k_k = 0  \qquad {\text{for all}}\ v \in \g \ .
\end{equation*}
Using the connectedness of $G$, this is the same as saying that $r$ is a special-unitary representation. 
In the much studied abelian linear sigma-model, which has $X = \CC^n$, $G = U(1)$ and
$r(\lambda) = {\rm diag} (\lambda^{q_1}, \ldots , \lambda^{q_n}  )$, 
the equivariant Calabi-Yau condition is thus just $\sum_k q_k = 0$, as found in \cite{W4}.  

$\ $

Our second example is a generalization of the abelian sigma-model. Let
\begin{equation*}
\begin{CD}
X = \bigoplus_k V_k  @>{\pi_X}>> M
\end{CD}
\end{equation*}
be a sum of holomorphic vector bundles over a complex manifold $M$. Then, after choosing a covariant derivative on 
$X \rightarrow M$, there is a natural isomorphism between the tangent bundle $TX \rightarrow X$ and the pull-back bundle
\begin{equation}
\pi_X^\ast (TM \oplus X) \longrightarrow X  \ . 
\label{2.14}
\end{equation}
Now let the circle $U(1)$ act on each $V_k$ by scalar multiplication
with charge $q_k$. This defines a global and holomorphic action of $U(1)$ on $X$.
This action, of course, lifts to $TX$, and under the isomorphism with (\ref{2.14}) the lift corresponds to the sum of 
the trivial action on $TM$ and the ``non-lifted'' action on $X$. The usual properties of Chern classes, which 
also hold in the equivariant case, then allow us to compute that
\begin{align*}
c_1^G (TX) &=  \pi_X^\ast \, c_1^G (TM\oplus X) = \pi_X^\ast \, [ c_1 (TM) + \sum_k c_1^G (V_k) ] \\
&=   \pi_X^\ast \Bigl\{ c_1 (TM) + \sum_k[ c_1 (V_k)  - e^1 \otimes q_k (\text{rank }V_k)  /(2\pi)] \Bigr\}  \ ,
\end{align*}
where $e^1$ is the single generator of the Lie algebra ${\mathfrak u}(1)$. Thus the manifold $X$ with this action 
is topologically an equivariant Calabi-Yau if and only if
\begin{equation*}
\begin{cases}
\sum_k q_k (\text{rank }V_k) = 0  \\
c_1 (TM) + \sum_k c_1(V_k) = 0 \ .
\end{cases}
\end{equation*}
Observe that when $M$ is a Riemann surface the second equation is just the numerical condition
$(2 - 2 g_M) + \sum_k \text{deg}\ V_k  = 0$. This agrees with \cite{B-P}, where these equivariant Calabi-Yau's
were constructed for $M$ a Riemann surface and $X\rightarrow M$ the sum of two line bundles.

$\ $

Finally, for the third example\footnote{This came up in a conversation with Andriy Haydys.}, let $(X, g)$ be a 
$4n$-dimensional hyperk\"ahler manifold with complex structures
$I$, $J$ and $K$, and associated K\"ahler forms $\omega_1$, $\omega_2$ and $\omega_3$. It is then well known that the
combination $\omega := \omega_2 + i \omega_3$ is a closed and non-degenerate 2-form on $X$ that is holomorphic with
respect to the complex structure $I$ \cite{H-K-L-R}. In particular this implies that the wedge product $\Omega := \omega^n$
is a trivialization of the canonical bundle of $(X, I)$. Now, if $X$ is also equipped with a $G$-action that preserves 
the hyperk\"ahler structure, then it is clear that $\Omega$ will be $G$-invariant, or in other words $X$ will be 
$G$-equivariantly Calabi-Yau. Moreover, if the $G$-action on $X$ is tri-hamiltonian, i.e. if there exists a hyperk\"ahler 
moment map $(\mu_1, \mu_2, \mu_3): X \rightarrow {\mathbb{R}}^3 \otimes \g^\ast$, then by definition the action on 
$(X, I, \omega_1)$ is hamiltonian with moment map $\mu_1$. All this, of course, would certainly be expectable, as the 
hyperk\"ahler condition is stronger than the Calabi-Yau one, and so compatibility of the $G$-action with the hyperk\"ahler 
structure naturally entails compatibility with the Calabi-Yau structure. The advantage here is that there already exists
a good pool of non-trivial examples of hyperk\"ahler manifolds with compatible $G$-actions, both in the abelian and 
non-abelian cases, and so we obtain for free examples of equivariant Calabi-Yau's. We list below a few of the most famous 
among these tri-hamiltonian hyperk\"ahler manifolds.

{\bf{(i)}} The well known Taub-NUT and gravitational multi-instanton spaces, as well as the Calabi spaces $T^\ast {\mathbb{CP}}^n$, all possess hyperk\"ahler
structures invariant under the action of at least circles (see \cite{G-R-G}).

{\bf{(ii)}} The toric hyperk\"ahler manifolds of \cite{Biel-Dan} are all equipped with tri-hamiltonian actions of the torus
 $T^n$, where $4n$ is the real dimension of the manifold.

{\bf{(iii)}} Let $G$ be a compact Lie group and $G^\CC$ its complexification. Then the cotagent bundle $T^\ast G^\CC$
carries a natural hyperk\"ahler structure that is invariant with respect to the $G\times G$-action induced by the left and
right translations on the group. This hyperk\"ahler structure is defined through the identification of $X=T^\ast G^\CC$ with
the space of solutions of Nahm's equations on the closed interval $[0, 1]$, modulo gauge transformations that are fixed
at the boundary of the interval \cite{Kron1}. 

{\bf{(iv)}} Assume that the compact group $G$ is semi-simple, and let $T$ be a maximal subtorus. Then the quotient
$G^\CC / T^\CC$ also carries hyperk\"ahler structures that are invariant under the natural $G$-action on this space. 
These structures are obtained by identifying $X= G^\CC / T^\CC$ with the moduli space of certain classes of instantons 
over ${\mathbb{R}}^4 \! \setminus \! \{0 \}$ \cite{Kron2}. 

{\bf{(v)}} Let $(S, g)$ be a 3-Sasakian manifold acted by a compact connected group $G$ of 3-Sasakian isometries. Then the
cone $C(S):= {\mathbb{R}^+} \times S$ with metric $\bar{g} = \dd t^2 + t^2 g$ has a natural hyperk\"ahler structure 
which is invariant by the trivial extension to $C(S)$ of the $G$-action on $S$ \cite{B-G-M}.

\subsection{Twisting}

Twisting a ${\mathcal N} =(2,2)$ supersymmetric theory is a very standard procedure; see \cite{W1, W2} for the original 
constructions and \cite{W3, Hori-al} for detailed reviews in the case of non-gauged sigma-models. Twisting is 
performed along the non-anomalous R-symmetries of the theory, and so for a general K\"ahler target 
$X$ there is only one twist, the A-twist, performed along the vector R-symmetry; if $X$ is in addition 
equivariantly Calabi-Yau, i.e. $c_1^G (TX) = 0$, then twisting along the axial symmetry provides a 
second topological theory, the B-theory.

Twisting, in practice, leads to a reinterpretation of the fields of the supersymmetric theory such that
the lagrangian makes sense on any Riemann surface $\Sigma$, not just the flat $\Sigma$'s of the SUSY 
theory; this reinterpretation is done according to precise rules and at the end, for example, all the
spinor fields are regarded either as scalars or one-forms on $\Sigma$ (with values on vector bundles).
These precise rules are as follows. Each of the fields in (\ref{2.1}) is in a space of sections
$\Omega^0 (\Sigma ; L \otimes V)$, where $V$ can be $\phi^\ast (\ver)$, $\gp$ or $\gp^\CC$, and $L$ is either
$K^{\pm 1/2}$ or the trivial bundle $\CC$. On the other hand, each field of (\ref{2.1}) is acted by the vectorial
R-symmetry (\ref{2.5}) with charge $q_V$ and by the axial symmetry (\ref{2.6}) with charge $q_A$. The rules then say 
that, after the topological twist, the field in question should be regarded as a section of $L' \otimes V$, 
where $L' = L \otimes K^{q_V /2}$ for the A-twist and  $L' = L \otimes K^{q_A /2}$ for the B-twist. Applying
this rule to all the fields of the gauged sigma-model one gets the following nice little table, in the manner 
of \cite{Hori-al}, 
\begin{equation}
\begin{array}{|>{}c | >{}c >{}c   >{}c  | >{}c  |  >{}c  |}

\hline             &           &\text{SUSY}    &             &\text{A-twist}  &\text{B-twist}     \\
                   &  U(1)_V   &    U(1)_A     &    L        &     L'         &       L'          \\
\hline    
           \psim   &    -1     &       1       &   K^{1/2}    &    \CC        &      K            \\
           \psibm  &     1     &      -1       &   K^{1/2}    &    K          &      \CC          \\
           \psip   &    -1     &      -1       &   K^{-1/2}   &    K^{-1}      &      K^{-1}        \\
           \psibp  &     1     &       1       &   K^{-1/2}   &    \CC        &      \CC          \\ 
            F      &    -2     &       0       &   \CC       &    K^{-1}      &      \CC          \\ 
           \ov{F}  &     2     &       0       &   \CC       &     K          &      \CC          \\
           \lamm   &     1     &       1       &   K^{1/2}    &     K          &       K           \\
           \lambm  &    -1     &      -1       &   K^{1/2}    &     \CC        &      \CC          \\  
           \lamp   &     1     &      -1       &   K^{-1/2}   &    \CC         &      K^{-1}       \\
           \lambp  &    -1     &       1       &   K^{-1/2}   &    K^{-1}       &      \CC          \\
           \si     &     0     &       2       &   \CC       &     \CC        &      K            \\
           \sib    &     0     &      -2       &   \CC       &     \CC        &      K^{-1}       \\
             D     &     0     &       0       &   \CC       &     \CC        &      \CC          \\
\hline
\end{array}
\label{2.11}
\end{equation}

In addition to the ``reinterpreted'' fields, each twisted theory is endowed with a fermionic operator
whose action on the fields is just a particular combination of the supersymmetry transformations 
(\ref{2.3}) and (\ref{2.4}).  More explicitly, and following the convention of \cite{Hori-al}, define the operators $Q_\pm$
and $\ov{Q}_\pm$ by
\begin{equation}
\delta = \epsilonp Q_- - \epsilonm Q_+ - \epsilonbp \ov{Q}_- + \epsilonbm \ov{Q}_+  \ ,
\label{2.12}
\end{equation}
where $\delta$ is given by (\ref{2.3}) and (\ref{2.4}). Then the fermionic operator of the A-twist is defined as
$Q_A = Q_- + \ov{Q}_+$ and the operator of the B-model as $Q_B = \ov{Q}_- + \ov{Q}_+$.

\section{The gauged A-twist}

\subsection{Fields, action and the $Q_A$-operator}

Proceeding impartially by alphabetical order, we start with the A-model. 
Define formally a new set of fields by the formulae:
\begin{align}
\chi^k &= \sq \psim^k    &      \psi_z^a &= -i \lamm^a     \label{3.1}  \\
\ov{\chi^k} &=  \sq \, \ov{\psip^k}     &      \psi^a_{\zb}  &=  i  \ov{\lamp^a}   \nonumber \\
\varphi^a  &=  -2\sq \, i  \si^a     &      \rho_{\zb}^k  &= \sq \, \psip^k      \nonumber \\
\la^a  &=   \sib^a / (2\sq)   &      \ov{\rho^k_{\zb}}  &=  \sq\, \ov{\psim^k}    \nonumber \\
\eta^a &=  (\ov{\lamm^a} + \lamp^a)/2i   &    c^a &= i (\ov{\lamm^a} - \lamp^a)    \nonumber \\
H^k_{\zb}  &=   4i\dd^A_{\zb} \phi^k + 2(F^k - \Gamma^k_{ij} \psip^i \psim^j  )  &     C^a &= 2 (F_A)^a_{12} + 2 D^a   \ . \nonumber 
\end{align}
The interpretation of the new fields as scalars or 1-forms comes, as explained before, from table (\ref{2.11}). 
These local components can be combined to define the global fields
\begin{align*}
\chi  &\in  \Omega_-^0 (\Sigma ; \phi^\ast \ver )  &      \varphi, \la , C  &\in \Omega_+^0 (\Sigma ; \gp )     \\
\rho  &\in  \Omega_-^{0,1} (\Sigma ; \phi^\ast \ver )  &   \eta , c  &\in \Omega_-^0 (\Sigma ; \gp )         \\
H  &\in  \Omega_+^{0,1} (\Sigma ; \phi^\ast \ver )  &      \psi   &\in \Omega_-^1 (\Sigma ; \gp )  \ .
\label{}
\end{align*} 
The other ``overlined'' fields are then to be interpreted as the local complex conjugates of these ones.

The action of the fermionic operator $Q_A = Q_- + \ov{Q}_+$ on the new fields follows from the supersymmetry 
transformations (\ref{2.3}), (\ref{2.4}) and the definition (\ref{2.12}) of $Q_-$ and $\ov{Q}_+$. In fact, one simply needs to substitute 
the new fields (\ref{3.1}) into the supersymmetry transformations, put $\epsilonp = \epsilonbm = 1 $ and 
$\epsilonm = \epsilonbp = 0$, and finally write the result in an invariant form that 
makes sense on any Riemann surface $\Sigma$. This procedure yields:
\begin{align}
Q_A\, \phi^k   &=  \chi^k     &     Q_A \, A  &=  \psi     \label{3.3}   \\
Q_A\, \chi^k  &=  \varphi^a \he_a^k     &   Q_A \,\psi &=  -  \nabla^A \varphi    \nonumber  \\
Q_A\, \la &= \eta     &   Q_A\, c &= C     \nonumber  \\
Q_A\, \eta &= [\varphi , \la]     &     Q_A\, C   &=  [\varphi, c]     \nonumber \\
Q_A\, \rho^k  &=  H^k  -  \Gamma^k_{ij} \chi^i  \rho^j   &   Q_A\, \varphi  &=  0   \nonumber   \\
Q_A\, H^k  &=  R_{i\bar{j}l\bar{m}} h^{k\bar{j}}  \chi^l  \ov{\chi^m} \rho^i  -   \Gamma^k_{jl} H^j \chi^l
 +  \varphi^a (\nabla_j \he_a^k)  \rho^j    \ . \nonumber 
\end{align}
The apparently random numerical factors in (\ref{3.1}) were chosen such as to render these last transformations as simple as possible.
The result also agrees with \cite{Bap}, modulo the notations.

The topological action of the A-theory is also obtained by simple substitution of the new fields into the supersymmetric 
lagrangian (\ref{2.2}). The result, including the auxiliary fields, is
\begin{equation*}
\begin{split}
I_A  =    \int_\Sigma & \Bigl\{  \frac{1}{2e^2} |F_A|^2  + |\dd^A \phi|^2   + 2e^2 |\mu \circ \phi|^2  
+ \frac{i}{e^2} \langle \nabla^A \varphi , \nabla^A \la  \rangle   +  \frac{1}{2e^2} |[\varphi , \la]|^2   \\ 
&+ \frac{1}{2e^2} [\varphi, \eta]_a \eta^a   +  \frac{1}{8e^2} [\varphi , c]_a c^a  
- \frac{1}{2e^2}  |\frac{1}{2} C  - \ast F_A - 2e^2 \: \mu \circ \phi |^2     \\
&-  \frac{1}{8} |H - 4i \db^A \phi |^2  
+ i h_{j\bar{k}} (\varphi^a \la^b + \varphi^b \la^a ) \he_a^j \, \ov{\he_b^k}  
+ 2 i h_{j\bar{k}} (\nabla_l \he_a^j ) \la^a \chi^l \ov{\chi^k}   \\  
&+ i h_{j\bar{k}} (\eta^a + \frac{1}{2} c^a )\, \ov{\he^k_a}\, \chi^j  
+ i h_{j\bar{k}} (\eta^a - \frac{1}{2} c^a )\, \he^j_a \, \ov{\chi^k}  \,  \Bigr\} \,\, \rm{vol}_\Sigma   \\
&+ \frac{i}{e^2}  \eta_a  \nabla^A \ast \psi^a - \frac{1}{2e^2} c_a  \nabla^A \psi^a 
+ \frac{i}{8}  R_{i\bar{j}k \bar{m}}  (\rho^i \wedge \ov{\rho^j} ) \chi^k  \ov{\chi^m}  \\
&- \frac{i}{e^2} \la_a [\psi, \ast \psi]^a  
+ \frac{1}{2} h_{j\bar{k}}\,  \rho^j \wedge (\phi^\ast \nabla^A ) \ov{\chi^k}
+ \frac{1}{2} h_{j\bar{k}} \, \ov{\rho^k} \wedge (\phi^\ast \nabla^A ) \chi^j  \\
&+ \frac{i}{8} h_{j\bar{k}}  \varphi^a (\nabla_l \he^j) \rho^l \wedge \ov{\rho^k}
+ \frac{1}{2} h_{j\bar{k}}\,  \he_a^j \, \psi^a  \wedge \ov{\rho^k}
+ \frac{1}{2} h_{j\bar{k}} \, \ov{\he_a^k}\, \psi^a  \wedge \rho^j  \ .
\end{split}
\label{3.4}
\end{equation*}
This action is $Q_A$-exact up to topological terms, just as in the non-gauged model of \cite{W2}. One can in fact 
check that
\begin{equation}
I_A \: =\:  Q_A  \Psi \ + \ \int_\Sigma \phi^\ast [\eta_E]
\label{3.5}
\end{equation}
with gauge fermion
\begin{equation*}
\begin{split}
\Psi \ = \ \int_\Sigma  \Bigl\{&  \frac{1}{2e^2}\, c_a (\ast F_A + 2e^2 \mu \circ \phi)^a  + \frac{1}{8e^2}\, c_a C^a 
+ \frac{1}{2e^2}\, \eta_a [\varphi , \la]^a  + i h_{j\bar{k}}\, \la^a (\he_a^j \,\ov{\chi^k} + \ov{\he_a^k}\, \chi^j )  
\Bigr\} \rm{vol}_\Sigma    \\
&+ \frac{i}{e^2}\, \la_a (\nabla^A \ast \psi^a)  - \frac{i}{16}\, h_{j\bar{k}} \,\ov{\rho^k} \wedge (H - 8i \db^A \phi)^j 
+ \frac{i}{16}\, h_{j\bar{k}}\, \rho^j \wedge \ov{(H - 8i \db^A \phi)^k} \ .
\end{split}
\end{equation*}
The topological term on the right-hand-side of (\ref{3.5}) can be described as follows.
The symbol $[\eta_E]$ represents a cohomology class in $H^2 (E)$. It is the class represented by the 2-form
\begin{equation*}
\eta_E (A) = \omega_X  - \dd (\mu_a A^a)  \qquad \in \Omega^2 (P \times X) \ ,
\end{equation*}
which descends to $E = P\times_G X$. This form is manifestly closed, for the K\"ahler form $\omega_X$ on $X$ is
closed, and its cohomology class does not to depend on $A$. 
It is also clear that 
$\int_\Sigma \phi^\ast [\eta_E]$ does not change under deformation of $\phi$, since the pull-back map is always
homotopy invariant, so this term is indeed topological.

 Finally, if desired, the auxiliary fields $C$ and $H$ can be eliminated from the action and the $Q_A$-transformations 
through their equations of motion
\begin{align*}
C^a &=   2 \ast F_A^a + 4e^2 \: \mu^a \circ \phi  \\
H^k_{\zb} &=  4i\dd^A_{\zb} \phi^k \ .
\end{align*}

One should also observe that the topological action $I_A$ is gauge invariant. The standard methods of local quantum 
field theory therefore recommend that it be gauge-fixed through the introduction of Fadeev-Popov ghost fields. This
can presumably be done as explained in \cite{Baul-Sin}, and would simply amount to adding to $I_A$ a further $Q_A$-exact 
term.

\subsection{Observables}

Having described the field content, the lagrangian and the $Q_A$-transformations of the theory, the next step is to 
look for an interesting set of observables whose correlation functions we would like to compute. In the non-gauged 
A-model the standard procedure is to construct such observables from the de Rham cohomology classes of the target
$X$. In the gauged model, of course, the analog procedure uses instead the $G$-equivariant cohomology classes of 
$X$. This construction was first described in \cite{W2}, and then with a little more detail in \cite{Bap}.

Recall that the $G$-equivariant complex $\Omega^\bullet_G (X)$ is the set of $G$-invariant elements in the tensor product 
$S^\bullet (\g^\ast) \otimes \Omega^\bullet (X)$.  A typical equivariant form $\alpha $ may thus be locally written as    
\begin{equation*}
\alpha \ = \ \alpha_{a_1 \cdots a_r k_1 \cdots k_p \bar{l_1} \cdots \bar{l_q}}
(w) \ \zeta^{a_1} \cdots \zeta^{a_k} \; \dd w^{k_1} \wedge \cdots \wedge 
\dd w^{k_p} \wedge \dd \bar{w}^{l_1} \wedge \cdots \wedge  \dd \bar{w}^{l_q} \ ,
\label{}
\end{equation*}
where  the coefficients $\alpha_{a_1 \cdots a_r k_1 \cdots k_p \bar{l_1} \cdots \bar{l_q}}$
are symmetric on the $a_j$'s and anti-symmetric on the $k_j$'s and $l_j$'s. To each such form one can associate 
an operator $\OO_\alpha$ in the topological field theory defined by the local formula
\begin{equation}
\OO_\alpha  \: = \: (\alpha_{a_1 \cdots a_r k_1 \cdots k_p \bar{l_1} \cdots \bar{l_q}}
\circ \phi )  \left[\prod_{j=1}^r    (\varphi + \psi + F_A
)^{a_j} \right]  \left[ \prod_{i=1}^p  (\chi^{k_i} + \dd^A \phi^{k_i} ) \right]  
\left[ \prod_{i=1}^q  (\ov{\chi^{l_i}} + \dd^A \ov{\phi^{l_i}} ) \right]  \, .
\label{3.6}
\end{equation}
It can then be checked that this correspondence is globally well defined and that, furthermore,
\begin{equation}
(\dd_{\Sigma} + Q_A ) \; \OO_\alpha  =   \OO_{\dd_G \alpha}  \ ,
\label{3.7}
\end{equation}
where $\dd_\Sigma$ is the exterior derivative on $\Sigma$ and $\dd_G$ is the Cartan operator on $\Omega^\bullet_G (X)$. Now assume that $\alpha$ is 
$\dd_G$-closed and decompose $\OO_\alpha$ according to the form degree over $\Sigma$,
i.e. write 
\begin{equation*}
\OO_\alpha \ = \ \OO_\alpha^{(0)} \ +\ \OO_\alpha^{(1)} \ +\ \OO_\alpha^{(2)}\ ,  
\end{equation*}
where for example
\begin{equation}
\OO_\alpha^{(0)}  \ = \ (\alpha_{a_1 \cdots a_r k_1 \cdots k_p \bar{l_1} \cdots \bar{l_q}}
\circ \phi )  \left(\prod_{j=1}^r   \varphi^{a_j} \right)  \left( \prod_{i=1}^p  \chi^{k_i}  \right)   
\left( \prod_{i=1}^q  \ov{\chi^{l_i}} \right)  \ .
\label{3.8}
\end{equation}
Then in terms of this decomposition identity (\ref{3.7}) breaks into
\begin{align*}
\dd_\Sigma \ \OO_\alpha^{(2)} \ &= \ 0 \ ,  \\
\dd_\Sigma \ \OO_\alpha^{(1)} \ &= \ - \; Q_A \; \OO_\alpha^{(2)} \ ,   \\
\dd_\Sigma \ \OO_\alpha^{(0)} \ &= \ - \; Q_A \; \OO_\alpha^{(1)} \ ,   \\
Q_A \; \OO_\alpha^{(0)} \ &= \ 0 \ ,
\label{3.9}
\end{align*}
which are the descent equations of the model. Finally let $\gamma$ be any $j$-dimensional homology cycle in 
$\Sigma$ and define the new operators 
\begin{equation*}
W (\alpha , \gamma ) \ := \ \int_\gamma \ \OO_\alpha^{(j)}  \ .  
\label{3.10}
\end{equation*}
These are then the natural observables associated with the gauged A-model. In fact it follows as usual from  
the descent equations and Stokes' theorem that $W (\alpha, \gamma )$ is $Q_A$-closed, so is indeed an observable. 
Moreover, the $Q_A$-cohomology class of $W (\alpha , \gamma )$ only depends on the classes of $\alpha$ and 
$\gamma$ in $H^\bullet_G (X)$ and $H_j (M)$, respectively.
The typical correlation functions of the theory can then be written down as path-integrals of the form
\begin{equation}
\int {\mathcal D} (A, \phi, \varphi,  \xi,\rho , \eta , c,  \psi ,  \chi) \ \ 
e^{-I_{A}} \ \prod_i \: W(\alpha_i , \gamma_i ) \ ,
\label{3.11}
\end{equation}
where the integration is taken over all fields, but with $\phi$ restricted to a fixed topological sector,
or more precisely with fixed class $\phi_{\ast} [\Sigma] \in H_2^G (X)$.

\subsection{Localization and moduli space}

The usual credo says that a path-integral with a fermionic symmetry localizes to the bosonic field 
configurations that are fixed points of the symmetry. Since $Q_A$ can be regarded as a generator of
one such  symmetry, we will be interested in the bosonic field configurations annihilated by $Q_A$. These
field configurations can be read from (\ref{3.3}) and, after eliminating the auxiliary fields, are precisely 
the solutions of
\begin{align}
&\db^A \phi = 0          \label{3.12}   \\
&\ast F_A  + 2 e^2 \mu \circ \phi = 0    \nonumber  \\
&\nabla^A \varphi = \varphi^a  (\he_a \circ \phi) = 0  \ . \nonumber 
\end{align}  
The first two equations are known as the general vortex equations on a Riemann surface. They were first
written down in \cite{C-G-S;MiR} and generalize the usual Nielsen-Olsen vortex equations. The two equations
involving $\varphi$, although in general non-trivial, in many cases of interest only have the 
$\varphi = 0$ solution, and so in these cases can be discarded. It can be shown, for example, that if
$0$ is a regular value of the moment map $\mu$, then given any fixed homotopy class of sections of $E$, 
for a sufficiently big value of the constant $e^2\, (\text{Vol }\Sigma)$ any solution of (\ref{3.12}) with $\phi$
in that class has zero $\varphi$ \cite[lem. 4.2]{C-G-M-S}. Another instance, in the abelian case: if $G$ is
a torus, $X$ is compact connected and $(\int_\Sigma F_A)/ (e^2 \text{Vol }\Sigma)$ is a regular value 
of $\mu$, then any solution of (\ref{3.12}) has zero $\varphi$ \cite{Bap}.
Nonetheless, even after discarding the last line of (\ref{3.12}), the two remaining (vortex) equations are very non-trivial.
For example, unlike monopoles or instantons, no explicit non-trivial solution of these equations is known, and this for any $\Sigma$, 
$X$ or $G$, including the non-compact  $\Sigma =\CC$. 

For the topological field theory, however,
the main objects of interest are not the solutions themselves, but rather the spaces of all solutions,
or more precisely the moduli spaces of solutions up to gauge equivalence. These vortex moduli spaces are
in general finite-dimensional, have a natural K\"ahler structure, but may contain singularities and be
non-compact. Their virtual complex dimension, as given by elliptic theory, is
\begin{equation}
({\rm dim}_\CC X - {\rm dim}G )(1-g) +   \langle  c_1^G (TX)\: ,\: \phi (\Sigma) \rangle \ ,
\label{3.13}
\end{equation}
and is basically just the difference of the indices of the operators in (\ref{2.7}) \cite{C-G-M-S}.

The standard heuristic arguments of TFT \cite{W1, W2} then say that, in favourable cases, the path-integrals 
(\ref{3.11}) reduce to finite-dimensional integrals of differential forms over the vortex moduli spaces. These
finite-dimensional integrals are completely classical objects and, modulo (in fact very difficult) problems
related to the singularities and non-compactness of the moduli spaces, make sense in the realm of traditional 
mathematics, as opposed to the path-integrals. The numbers provided be these finite-dimensional integrals 
can in fact be identified with the so-called Hamiltonian Gromov-Witten invariants of $X$, which have been
defined using a very different, rigourous, universal construction. All this story is analogous to the well 
known case of the non-gauged sigma-model, which leads to the Gromov-Witten invariants; it is spelled out
in detail in \cite{Bap}.

Another important fact is that in the limit $e^2 \rightarrow +\infty$ the gauged sigma-model with target $X$
tends to a non-gauged sigma-model with target $X/\!/G$. This is just as in the linear case of \cite{W4}. As a 
consequence one expects some relation to exist between the HGW-invariants of $X$ and the GW-invariants of 
$X/\!/G$ \cite{Ga-Sa}.

 We now end this section with a few references. Regarding the vortex moduli spaces, there has been a longstanding 
interest in them. Starting with the simplest case of the abelian Higgs models --- where $X = \CC$ and 
$G = U(1)$ --- about thirty years ago, the structure of these spaces has been investigated in several 
particular examples, mainly with $X$ a vector space. A hectic set of references is for example
\cite{Ba} within the more mathematical literature and \cite{E-I-N-O-S, W4, M-P} within theoretical physics. 
The Hamiltonian Gromov-Witten invariants, in comparison, have only recently been defined \cite{C-G-M-S, C-G-S;MiR}. They have been 
furthermore studied in \cite{C-S, Ga-Sa}.

%
%
%
%
%

\section{The gauged B-twist and Landau-Ginzburg models}

\subsection{Fields, action and the $Q_B$-operator}

Starting with the supersymmetric model of section 2, keep the fields $A$, $\phi$ and $D$ unchanged and, with the others, 
define formally a new set of fields through the expressions
\begin{align}
\rho^k_z  &=  \sq i \psim^k     &           \ov{\eta^k}  &= \sq (\ov{\psip^k} + \ov{\psim^k})     \label{4.1}   \\
\rho^k_{\zb}  &=  - \sq i \psip^k     &     \theta_k  &=  \sq \, h_{k\bar{j}} (\ov{\psim^j} - \ov{\psip^j})     \nonumber   \\
\xi^a_z   &=  - i \si^a / \sq   &      \psi_z^a  &=  \lamm^a                \nonumber    \\
\xi^a_{\zb}   &=   i \sib^a / \sq    &      \psi_{\zb}^a  &=  \lamp^a              \nonumber      \\ 
\vp^a  &=   \frac{i}{2} (\lambm^a + \lambp^a )  &   \mathcal{F}^k  &=  F^k \, -\,  \Gamma^k_{ij}\, \psip^i \psim^j     \nonumber \\
\lambda^a  &=   \frac{i}{2} (\lambm^a - \lambp^a )  &   \ov{\mathcal{F}^k}  &=  \ov{F^k} \, -\, \ov{\Gamma^k_{ij}}\, \,
\ov{\psim^i}\, \, \ov{\psip^j}  \ . \nonumber 
\end{align}
These local components can be combined to define the global fields
\begin{align*}
\rho    &\in  \Omega_-^1 (\Sigma ; \phi^\ast \ver )  &           \xi   &\in  \Omega_+^1 (\Sigma ; \gp )     \\
\eta    &\in  \Omega_-^0 (\Sigma ; \phi^\ast \ver )  &           \vp , \lambda  &\in  \Omega_-^0 (\Sigma ; \gp )   \\ 
\theta  &\in  \Omega_-^0 (\Sigma ; \phi^\ast (\ver)^\ast  )  &   \psi   &\in  \Omega_-^1 (\Sigma ; \gp )           \\
{\mathcal F}\ &\in  \Omega_+^0 (\Sigma ; \phi^\ast \ver )    &   D  &\in  \Omega_+^0 (\Sigma ; \gp )  \ .
\label{}
\end{align*}
These latter fields, together with $A$, $\phi$ and $D$, form the field content of the gauged B-model.

The action of the fermionic operator $Q_B = \ov{Q}_+ + \ov{Q}_-$  follows from the supersymmetry transformations
(\ref{2.3}),(\ref{2.4}) and the definition (\ref{2.12}) of $\ov{Q}_\pm$. One simply needs to substitute the new fields (\ref{4.1}) into 
the supersymmetry 
transformations and then write the result in an invariant form that makes sense on any Riemann surface $\Sigma$. 
This procedure yields:
\begin{align}
Q_B \, \phi^k  &= 0     &    Q_B\, \ov{\eta^k} &= 0      \label{4.3}  \\
Q_B\, \ov{\phi^k} &=  \ov{\eta^k}     &   Q_B\, \theta_k    &=  4 h_{j\bar{k}}\, \ov{\mathcal{F}^k}    \nonumber \\
Q_B \,\rho  &=  4 \dd^{A+i\xi} \phi    &    Q_B \,\ov{\mathcal{F}^k}  &=  - \ov{\Gamma^k_{rl}}\, \ov{\mathcal{F}^r}\, \ov{\eta^l} \nonumber  \\
Q_B\, A  &=  -i \psi          &           Q_B\, \xi  &=  \psi                 \nonumber   \\
Q_B \,\lambda &= \ast \nabla^A \ast \xi  +  D   &       Q_B \,\psi  &=  0         \nonumber \\
Q_B\, \vp   &=  \ast \bigl( F_A - \frac{1}{2} \, [\xi , \xi] + i  \nabla^A \xi  \bigr) &  
Q_B\, D &= - \ast \nabla^{A+ i\xi} \ast \psi         \nonumber  \\
Q_B \,\mathcal{F}^k  &=  i \ast (\phi^\ast \nabla^{A+i\xi}) \rho^k + \frac{i}{8} (\partial_{\bar{s}} \Gamma^k_{ij}) \ov{\eta^s} 
\ast (\rho^i \wedge \rho^j ) - 4 i \, \vp^a \he_a^k      \ ,    \nonumber 
\end{align}
where $\ast$ is the Hodge operator on $\Sigma$. Observe that the complex connection
\begin{equation*}
{\mathcal A} = A + i \xi
\end{equation*}
emerges naturally in these transformations. It is a $Q_B$-closed field and has curvature
\begin{equation*}
F_{\mathcal A} = F_A  -  \frac{1}{2} [\xi, \xi]   + i \nabla^A \xi \ .
\end{equation*}

The topological action of the B-theory is also obtained by simple substitution of the new fields into the supersymmetric 
lagrangian (\ref{2.2}). After discarding a total derivative term $\dd [i \si_a (\dd^A \sib^a) /2 ]$ in that lagrangian the result is 
\begin{equation*}
\begin{split}
I_B =  \int_\Sigma  &\Bigl\{  \frac{1}{2e^2} |F_A - \frac{1}{2} [\xi , \xi] |^2 + |\dd^A \phi|^2 + 2 e^2 |\mu \circ \phi|^2 + |\xi^a \he_a |^2   
 + \frac{1}{2 e^2} |\nabla^A \xi|^2  \\ 
&+  \frac{1}{2e^2} |\nabla^A \ast \xi |^2 - \frac{1}{2 e^2} |D - 2 e^2 \mu \circ \phi |^2 
+ i \he_a^j (\vp^a \theta_j - h_{j\bar{k}} \lambda^a \ov{\eta^k})  \\
& - |\mathcal{F} +  \frac{1}{2}\,  {\rm grad}_\CC W |^2 
+  \frac{1}{4} h^{j\bar{k}} (\partial_j W)(\ov{\partial_k W}) 
+ \frac{1}{8}  h^{r\bar{k}}\, \theta_r \ov{\eta^j}\,  (\nabla_{\bar{k}} \partial_{\bar{j}} \ov{W})    \Bigr\} \rm{vol}_\Sigma    \\
&- \frac{1}{4} h_{j\bar{k}}\, \ov{\eta^k} (\phi^\ast \nabla^{A-i\xi}) \ast \rho^j  +  \frac{i}{4} \theta_j (\phi^\ast \nabla^{A+i\xi}) \rho^j 
- \frac{i}{32} (\partial_{\bar{s}} \Gamma^n_{jk}) \ov{\eta^s}\, \theta_n \, \rho^j \wedge \rho^k    \\
&- \frac{i}{2} h_{j\bar{k}}\, \ov{\he_a^k}\, \psi^a \wedge \ast \rho^j  + \frac{i}{e^2} \vp_a (\nabla^{A-i\xi} \psi^a )
- \frac{1}{e^2} \lambda_a (\nabla^{A-i\xi} \ast \psi^a )  -  \frac{i}{16} (\rho^j \wedge \rho^k )  (\nabla_k \partial_j W)  \ . 
\end{split}
\label{4.4}
\end{equation*}
When the superpotential $W$ is taken to be zero this action is $Q_B$-exact, just as in the usual non-gauged case \cite{L-L, W3}. In fact
after a few integrations by parts on can check that
\begin{equation*}
I_B \ = \ Q_B \: \Psi
\end{equation*}
with gauge fermion
\begin{equation*}
\begin{split} 
\Psi \: =\:  \int_\Sigma & \frac{1}{2e^2}\, \vp_a \, F_{A-i\xi}^a   
-  \frac{1}{4} h_{j\bar{k}} (\ast \rho^j)\wedge \ov{\dd^{A+i\xi} \phi^k}     - \frac{1}{4} {\mathcal{F}}^j \theta_j\,\rm{vol}_\Sigma      \\
&+ \frac{1}{2e^2} \lambda^a (\ast \nabla^a \ast \xi_a + 4 e^2 \mu_a \circ \phi - D_a)\: \rm{vol}_\Sigma   \ .
\end{split}
\end{equation*}
If desired, the auxiliary fields ${\mathcal F}$ and $D$ can be eliminated from the action and the $Q_B$-transformations 
through their equations of motion
\begin{align*}
\mathcal{F}^k &= - \frac{1}{2} h^{k \bar{l}} \partial_{\bar{l}} \ov{W}  \\
D^a &= 2 e^2 \mu^a \circ \phi  \ .
\label{}
\end{align*}

\subsection{Localization, moduli spaces and observables}

\subsubsection*{Localization}

As can be read from (\ref{4.3}), after eliminating the auxiliary fields the fixed points of $Q_B$ are 
the bosonic field configurations that satisfy
\begin{align}
&\dd^{\mathcal A} \phi = 0           &      F_{\mathcal A}  =  0     \label{4.6}  \\
&\ast \nabla^{\mathcal A} \ast \xi  + 2 e^2 \mu \circ \phi = 0    &   (\text{grad}_\CC W)\circ \phi = 0 \ . \nonumber 
\end{align}
Accordingly, one expects the path-integrals to localize to these configurations. Now, were this 
the A-model or the non-gauged B-model, nothing of major import would need to be added; in the
present case of the gauged B-model, however, there is one extra subtlety (already noted in 
\cite{W4} in the linear case) that allows us to take the localization argument a bit further. To explain this
 start by recalling that the operator $Q_B$ is defined as $\ov{Q}_+ + \ov{Q}_-$, 
where each of these two operators is defined through (\ref{2.12}) and makes perfect sense when acting
on the B-model fields (\ref{4.1}) defined on any Riemann surface. The first point to note is then that  
 the action $I_B$ is not only $Q_B$-exact, but also, up to topological terms, $\ov{Q}_+$-
and $\ov{Q}_-$-exact. One can in fact check that
\begin{equation*}
I_B  \: =\:  \ov{Q}_\pm \Psi_\pm \: \pm \:  2 \int_\Sigma \phi^\ast  \eta(A) \ , 
\label{4.7}
\end{equation*}
where the last term is topological, as in (\ref{3.5}), and the gauge fermions 
are\footnote{The notation $\rho_{\zb / z}$ means that the first option, here ${\zb}$, is to be taken 
for $\Psi_+$, and the second for $\Psi_-$; similarly for the other fields.}
\begin{equation*}
\begin{split}
\Psi_\pm =  \int_\Sigma &\Bigl\{  h_{j\bar{k}}\, h^{z\zb} (\rho^j_{\zb /z} \dd^A_{z/ \zb}\ov{\phi^k}
 -i \rho^j_{z/\zb}\, \xi^a_{\zb /z}\, \ov{\he_a^k} )  + 2 (\lambda^a \pm \vp^a ) (\mu\circ\phi)_a   \\
&\mp \frac{1}{2} h_{j\bar{k}}\, {\mathcal{F}}^j \, \ov{Q}_{\mp} \ov{\phi^k}\, 
\pm \, \frac{1}{e^2}\,  \ov{Q}_{\mp} ( \vp^a \lambda_a )  \Bigr\}  \rm{vol}_\Sigma        \  .
\end{split}
\end{equation*}
These fermions are just the components of $\Psi$ that transform with different charges under the
axial symmetry (\ref{2.6}), so that $\Psi = (\Psi_+ + \Psi_-)/2$; one can also check that 
$\ov{Q}_+ \Psi_- = \ov{Q}_- \Psi_+ = 0$. Now, with an action that is both $\ov{Q}_+$-
and $\ov{Q}_-$-exact, the expectation values of $\ov{Q}_\pm$-closed operators (such as 
$G$-invariant holomorphic functions on $X$) will localize to the simultaneous fixed points of
$\ov{Q}_+$ and $\ov{Q}_-$. These field configurations are of course also $Q_B$ fixed points, 
since $Q_B = \ov{Q}_+ + \ov{Q}_-$, but the converse needs not be true. While in the A-model
and in the non-gauged models these two sets of fixed points do in fact coincide, and so we
do not need to care about all this, in the gauged B-model the simultaneous fixed points of
$\ov{Q}_+$ and $\ov{Q}_-$ are the solutions to the seven equations   
\begin{align}
\dd^A \phi &=   F_A - [\xi , \xi]/2  = 0         &         \mu \circ \phi = 0      \label{4.8}  \\
\nabla^A \xi &=  \nabla^A \ast \xi = \xi^a \he_a = 0   &     (\text{grad}_\CC W)\circ \phi  = 0  \ , \nonumber 
\end{align}
which a priori seem to be stronger than (\ref{4.6}). 

\subsubsection*{Moduli spaces}

 In this section we will determine (in easy cases) the moduli space ${\mathcal M}_B$ of solutions of 
equations (\ref{4.8}) up to gauge equivalence.
Since ${\mathcal M}_B$ is the localization locus of the path-integrals, it is of course a very important 
object in the B-model. 

The easiest situation to analyse occurs when the Riemann surface $\Sigma$  has genus zero, 
so is a sphere. In this case it is well known that 
\begin{equation*}
\text{dim }\{ \xi :  \nabla^A \xi = \nabla^A \ast \xi = 0   \} = 
\text{dim } H^1_A (S^2 ; \gp) = 0
\end{equation*}
for every connection $A$, so that equations (\ref{4.8}) imply that $\xi = 0$ and that $F_A = 0$.
But on a sphere there are no monodromies, and the only possible flat connection is the trivial
connection on the trivial principal $G$-bundle, up to gauge equivalence. This means that one can
find a gauge transformation such that $\dd^A \phi = \dd \phi = 0$, and hence $\phi$ is 
gauge-equivalent to a constant map to the subset $\mu^{-1} (0)$ of $X$. This gauge transformation,
however, is unique only up to multiplication by a constant in $G$, and so it is clear that for 
genus zero 
\begin{equation}
{\mathcal M}_B \simeq 
\begin{cases}
\emptyset   &  \text{if $P$ is non-trivial,}  \\
\mu^{-1} (0) / G =  X/\! / G  &  \text{if $P$ is trivial and } W =0 \ , \\
(\mu^{-1} (0)  \cap \text{Crit }W  ) / G    & \text{if $P$ is trivial and } W \neq 0 \ ,
\end{cases}
\label{4.9}
\end{equation}
where constant maps have been identified with their target point.

$\ $

Although a priori not so evident, this result is also valid for $\Sigma$ of any genus provided that 
we assume that $G$ acts freely on $\mu^{-1} (0)$, i.e. provided that the symplectic quotient 
$X/\!/ G$ is smooth. To justify this we will now make a short detour. Start by noticing that the local 
equation  
\begin{equation*}
\dd^A \phi =  \dd \phi  +   A^a  (\he_a \circ \phi)  =  0 
\end{equation*}
implies that the image of a solution $\phi$ is contained in a single $G$-orbit in $X$; more precisely,
there exists a point $q \in \mu^{-1} (0)$ such that the image of $\phi : \Sigma \rightarrow E$ is 
contained in the sub-bundle
\begin{equation*}
E_q =  \{ [p, q] \in  E =P\times_G X  :  p  \in P    \} \  \subset   E  \ . 
\end{equation*}
Observe also that $E_{g\cdot q} = E_q$ for any $g \in G$ and that, by the assumed triviality of
the stabilizer of $q$, the map
\begin{equation*}
f_q  :  P  \longrightarrow E_q \ , \quad p \mapsto [p,q]   
\end{equation*}
is actually a fibre-preserving diffeomorphism. It is then clear that 
$f^{-1}_q \circ \phi : \Sigma \rightarrow P$ is a global section, and so $P$ must be trivial. Now 
consider the connection $A$. As is well known, such an object induces splittings of the
tangent bundles 
\begin{align*}
TP  &=  H_A  \oplus \ker \dd \pi_P       &    T E_q &= {\mathcal H}_A  \oplus \ker \dd (\pi_{E}|_{E_q}) \\
TE  &=  {\mathcal H}_A  \oplus \ker \dd \pi_E
\end{align*}
into horizontal and vertical sub-bundles. In this picture the covariant derivative of $\phi$ is just the
composition
\begin{equation*}
\begin{CD}
\dd^A \phi : T\Sigma @>{\dd \phi}>> TE  @>{{\rm projection}}>>  \ker \dd \pi_E \ ,
\end{CD} 
\end{equation*}
and so $\dd^A \phi = 0$ means that the image of $\dd \phi$ is entirely contained in ${\mathcal H}_A$.
But by the very definition of ${\mathcal H}_A$ we have that $\dd f_q (H_A) = {\mathcal H}_A$, which implies
that $f^{-1}_q \circ \phi$ is in fact a horizontal section of $P$, and this in turn shows that $A$ is 
gauge-equivalent to the trivial connection. From here onwards the same arguments as in the 
$\Sigma = S^2$ case lead to the conclusion that the moduli space ${\mathcal M}_B$ is given by (\ref{4.9}).

$\ $

The cases where $G$ does not act freely on $\mu^{-1} (0)$ are of course more complicated and 
difficult to analyse. Among these, the simplest situation occurs when $G$ acts freely everywhere in
$\mu^{-1} (0)$ except at $k$ fixed points. In this case, calling ${\mathcal C}_{\Sigma , P}$ the 
moduli space of solutions of 
\begin{equation*}
\nabla^A \xi =  \nabla^A \ast \xi = F_A - [\xi , \xi]/2  = 0 \ ,
\end{equation*}
it is rather clear that the space ${\mathcal M}_B$ will just consist of $k$ copies of 
${\mathcal C}_{\Sigma , P}$ when $P$ is non-trivial and, when $P$ is trivial, will be 
isomorphic to $X/ \! / G$ except that each singularity in this quotient (which corresponds to a fixed
point in $\mu^{-1} (0)$) is to be substituted by a copy of ${\mathcal C}_{\Sigma , P}$. Observe as well that in the
abelian case ${\mathcal C}_{\Sigma , P}$ is just
\begin{equation*}
{\mathcal C}_{\Sigma , P} \ \simeq \ H^1 (\Sigma)^{\dim G}  \times (\text{moduli space of flat 
connections on }P) \ .
\end{equation*}
These are of course only loose comments, and we will not pursue them here any further.

\subsubsection*{Observables}

The first natural observables of the B-theory are the holonomies, or Wilson loop operators, associated 
to the $Q_B$-closed complex connection ${\mathcal A}$. These observables, however, completely ignore the target
manifold $X$, and so if not coupled to other observables will have expectation values that only reflect properties
of the 2D-Yang-Mills. Another set of observables, this time dependent on $X$, are the $G$-invariant holomorphic 
functions on $X$. If $f$ is holomorphic on $X$ then the rules
\begin{align*}
Q_B \phi^k &= \ov{Q}_{\pm} \phi^k = 0       &    Q_B \theta_k &= 2 \ov{Q}_{\pm} \theta_k = -2\partial_k W
\end{align*}
show that $f\circ \phi$, besides being $Q_B$-closed, is $Q_B$-exact iff $f$ can be written as
\begin{equation}
f = v^k \partial_k W = \dd W (v)
\label{4.10}
\end{equation}
for some $G$-invariant holomorphic vector field $v$ on $X$. Thus the chiral ring of the gauged B-model is the ring of
$G$-invariant holomorphic functions on $X$ divided by the ideal of functions of the form (\ref{4.10}). All this is
analogous to the non-gauged sigma-model \cite{Hori-al}, one only has to add here the word $G$-invariant. Observe also that 
$G$-invariant holomorphic functions on $X$ descend to holomorphic functions on $X/ \! / G$, which, after localization,
is in some sense the ``effective target'' of the model. The author doesn't know, however, if every holomorphic 
function on $X/ \! / G$ can be obtained in this way, or more generally, how different is the $G$-invariant chiral
ring of $X$ from the standard chiral ring of $X/ \! / G$.

Finally, in the special case where the superpotential $W$ vanishes, a $G$-invariant form
\begin{equation*}
V \: = \: V_{\bar{i_1} \cdots  \bar{i_p}}^{j_1 \cdots  j_q} \: \: \dd \ov{w^{i_1}}\wedge \cdots \wedge \dd \ov{w^{i_p}}
\otimes  \frac{\partial}{\partial w^{j_1}} \wedge  \cdots \wedge  \frac{\partial}{\partial w^{j_q}}  
\qquad \in  \ \Omega^{0,p} (X; \Lambda^q TX)
\end{equation*}
determines an associated operator in the field theory by 
\begin{equation}
{\mathcal O}_V \: = \: V_{\bar{i_1} \cdots  \bar{i_p}}^{j_1 \cdots  j_q} \: \:  \ov{\eta^{i_1}} \cdots  \ov{\eta^{i_p}}\, 
 \theta_{j_1} \cdots  \theta_{j_q} \ .
\label{4.11}
\end{equation}
One can directly check that $ Q_B {\mathcal O}_V  = {\mathcal O}_{\db V}$, and so this correspondence defines a 
homomorphism between the $\db$-cohomology of $G$-invariant forms in $\Omega^{0,p} (X; \Lambda^q TX)$ 
and the $Q_B$-cohomology of operators in the B-model. Again,
all this mimicks the non-gauged model with the added $G$-invariant condition. Note, however, that (\ref{4.11}) is not
in general $\ov{Q}_{\pm}$-closed, and so more care is needed when localizing the expectation values of these  
observables, as explained at the beginning of section 4.2. This problem does not arise in the non-gauged B-model.

\vskip 25pt
\noindent
{\bf Acknowledgements.}
I would like to thank Guillaume Bossard, Lotte Hollands, Andriy Haydys, Sheer El-Showk and  David Tong for helpful conversations, 
as well as the JHEP referee for his comments. I am partially supported by the Netherlands Organisation for Scientific 
Research (NWO), Veni grant 639.031.616.

\appendix

\section{Notation and conventions}

\subsubsection*{Manifolds, group action and bundles}

 For reference, here is a list of the conventions and notation used in the paper.

$\ $

\noindent
$\bullet$ $\Sigma$ is a Riemann surface of genus $g$ and $X$ is a complex K\"ahler manifold.
$G$ is a compact connected Lie group that acts on $X$ on the left. The $G$-transformations preserve the symplectic and
complex structures of $X$. The Lie algebra of $G$ is called $\g$, has a basis $\{e_a \}$ and is equipped with an
Ad-invariant inner product $\kappa$, which may be used to identify $\g$ with the dual space $\g^\ast$. An element
$\xi \in \g$ induces a vector field $\hat{\xi}$ on $X$ whose flow is $p \mapsto \text{exp}(t\xi) \cdot p$. With this 
convention the Lie bracket on $\g$ is related to the Lie bracket of vector fields through 
$\widehat{[\xi_1 , \xi_2]} = - [\hat{\xi_1}, \hat{\xi_2}]$.

$\ $

\noindent
$\bullet$ The $G$-action on $X$ is assumed hamiltonian, i.e. there should exist a moment map $\mu : X \rightarrow \g^\ast$. In the 
convention used here the moment map satisfies
\begin{itemize}
\item[(i)]$\dd (\mu , \xi) = \iota_{\hat{\xi}}\, \omega_X$ for all $\xi \in \g$, where $\omega_X$ is the K\"ahler form on $X$
and $(\cdot , \cdot)$ is the natural pairing $\g^\ast \times \g  \rightarrow   {\mathbb R}$;  
\item[(ii)]$\rho_g^\ast \, \mu = {\rm Ad}_g^\ast \circ \mu $ for all $g
  \in G$, where $\rho$ denotes the $G$-action on $X$ and ${\rm Ad}^\ast $ is the coadjoint representation 
  on $\g^\ast$.
\end{itemize}
If a moment map $\mu$ exists, it is not in general unique, but all the
other moment maps have the form $\mu + r$, where $r \in [\g ,\g ]^0
\subset \g^\ast $ is a constant in the annihilator of $[\g, \g]$. Under the identification $\g^\ast \simeq \g$
provided by $\kappa$, the inner product, the annihilator $[\g, \g ]^0 $ is identified with the centre of $\g$. 
The constant $r$ is then the Fayet-Iliopoulos parameter of the supersymmetric theory.

$\ $

\noindent
$\bullet$ $\pi_P : P \rightarrow \Sigma$ is a principal $G$-bundle. $\pi_E : E \rightarrow \Sigma$ and $\g_P \rightarrow \Sigma$
are the associated bundles $E= P\times_G X$ and $\g_P = P\times_{\text{Ad}} \g$. These have typical fibres $X$ and $\g$,
respectively. The Higgs field $\phi : \Sigma \rightarrow E$ is a section of $E$. The vector bundle $\ver \rightarrow E$
is the kernel of the derivative $\dd \pi_E : TE \rightarrow T\Sigma$.

\subsubsection*{K\"ahler geometry}

Regarding the K\"ahler geometry of $\Sigma$ and $X$, we always work with the
holomorphic tangent bundles $T\Sigma$ and $TX$.
The local complex coordinates on $\Sigma$ and $X$ are $z= x^1 +i x^2$ and $\{ w^k \}$, respectively.
The hermitian metric $h_X$ is related to the real metric $g_X $ and the K\"ahler form $\omega_X $ by
\[
h \: =\:  h_{j \bar{k}}\,  \dd w^j  \otimes \dd \bar{w}^k \: =\:  g_X  - i\,  \omega_X \ .
\]
This implies that, with the most usual conventions for the wedge product, 
$\omega_X = (i/2) h_{j\bar{k}}\, \dd w^j \wedge \dd \bar{w}^k$. 
The hermitian (Levi-Civita) connection on $TX$ satisfies
\[
\nabla_{ \frac{\partial}{\partial w^j}} {\frac{\partial}{\partial
w^k}} \ = \ \Gamma_{ j k }^l \; \frac{\partial}{\partial
w^l} \ =\ h^{l \bar{r}}\, (\partial_j  h_{k
\bar{r}}) \;\frac{\partial}{\partial w^l}  \ .
\]
Its curvature components and Ricci form are then given by  
\begin{align*}
R_{j \bar{k} l \bar{r}} &=  - \partial_l \partial_{\bar{r}} h_{j \bar{k}}  +  h^{m \bar{n}}  (\partial_{l}  h_{j \bar{n} } ) 
(\partial_{\bar{r}}  h_{m \bar{k}} ) \\
\rho  &= -i \, \partial \bar{\partial} \log (\det h) \ .
\end{align*}
For any $\xi \in \g$ one can check that the holomorphic and Killing vector field $\hat{\xi}$ satisfies
\begin{align*}
h_{j \bar{k}}\, \nabla_l \hat{\xi}^k  &\: =\:  - h_{l \bar{k}}\,  \ov{\nabla_j \hat{\xi}^k}    \\
2\, \partial_{\bar{k}} (\mu, \xi)  &\: =\:  i h_{j \bar{k}} \, \hat{\xi}^j  \ .
\end{align*}

The Hodge star operator on $\Sigma$ satisfies
\begin{align*}
\ast \omega_\Sigma &=  1   &          \ast \dd z &= -i \dd z     \\
\ast 1  &=  \omega_\Sigma   &         \ast \dd \zb &=  i \dd \zb  \ .
\end{align*}

In sections 3 and 4 we have often used that a connection $\nabla^A$ on some bundle $V \rightarrow \Sigma$ can be 
extended to an operator $\Omega^r (\Sigma ; V) \rightarrow \Omega^{r+1} (\Sigma ; V)$, so beyond its usual $r=0$ definition.
For instance if $\psi = \psi_z \dd z + \psi_{\zb} \dd \zb$ is a one-form with values on $V$ then $\nabla^A \psi = 
(\nabla^A_z \psi_{\zb} - \nabla_{\zb} \psi_z) \dd z \wedge \dd \zb$.

\subsubsection*{The ${\mathcal N}=2$  lagrangian and supersymmetry transformations}

In section 2 we spelled out the euclidean lagrangian and supersymmetry transformations for the ${\mathcal N}=2$ gauged non-linear
sigma-model in two dimensions. These formulae are related to their counterparts on Minkowski space-time through the substitutions
\begin{align*}
2 \dd^A_{\zb}  &\longleftrightarrow  \dd_1^A +  \dd^A_0    &          
2 \nabla^A_{{\zb}/z}  &\longleftrightarrow  \nabla_1^A \pm  \nabla^A_0   \\
2 \dd^A_{z}  &\longleftrightarrow  \dd_1^A -  \dd^A_0    &         
2 (\phi^\ast\nabla^A)_{{\zb} /z}  &\longleftrightarrow  (\phi^\ast \nabla^A)_1 \pm  (\phi^\ast \nabla^A)_0     \\
F_{12}   &\longleftrightarrow  i F_{01}
\end{align*}
and a global sign change. Here $(x^0 , x^1)$ are the Minkowski coordinates on $\Sigma = {\mathbb R}^{1,1}$ with signature $(-,+)$ and
$x^1 + i x^2 =z$ is the complex coordinate on the euclidean $\Sigma = \CC$. The Minkowski lagrangian is real, i.e. invariant under 
complex conjugation, while the euclidean lagrangian is not. The conventional rules for conjugating fermions in Minkowski signature
are $\ov{\ov{\lambda}} = \lambda$ and $\ov{\lambda_1 \lambda_2} = \ov{\lambda}_2 \, \ov{\lambda}_1$. In euclidean signature these rules
do not apply. In fact, the barred and unbarred euclidean fermionic fields should be regarded as independent \cite{D-F},
and in rigour should have been denoted by different letters in section 2.

The Minkowski version of the lagrangian and supersymmetry transformations of section 2 were obtained by dimensional reduction of the 
${\mathcal N}=1$ formulae in four dimensions presented in \cite{D-F}. Since the conventions of \cite{D-F} differ from the most commonly used 
in the physics literature we have adjusted the various $i$ and $\sqrt{2}$ factors so that, upon specialization to the gauged linear 
sigma-model, our formulae agree with \cite{W4, W5}.

This specialization to the linear sigma-model and group $G =U(n)$ should, nevertheless, be done with some care, since the physicists identify
the Lie algebra of $U(n)$ with the hermitian matrices while in mathematics the conventional identification is with the anti-hermitian matrices. 
In the physics convention a Lie algebra valued field  such as $\sigma = \sigma^a e_a$ is identified with a hermitian matrix 
$\tilde{\sigma}$; our complex conjugate field $\sib = \sib^a e_a$ becomes the hermitian conjugate matrix $\tilde{\sigma}^\dag$; the Lie 
brackets $[\si, \sib] = \si^a \sib^b [e_a , e_b]$ become, on the other hand, $i [\tilde{\sigma}, \tilde{\sigma}^\dag]$.
This implies that the covariant derivative $\nabla^A \si$ of (\ref{2.13}) becomes 
$\dd \tilde{\sigma} + i [\tilde{A}, \tilde{\sigma}]$. Finally, for the natural action of $G=U(n)$ on $\CC^n$, one can calculate that
the vector fields on $T\CC^n \simeq \CC^n$ become 
\begin{align*}
\si^a (\he_a \circ \phi) &\longrightarrow i \tilde{\sigma} \phi      &       \si^a  \nabla_k \he_a^j   &\longrightarrow i \tilde{\sigma}^j_k \\               
\sib^a (\he_a \circ \phi) &\longrightarrow i \tilde{\sigma}^\dag \phi \ .
\end{align*}
Systematically applying these substitutions to all the fields in the lagrangian of section 2 (rotated to Minkowski space) we get exactly
the lagrangian of \cite{W4, W5}. As for the supersymmetry transformations, they agree with all the expressions of \cite{W5} except that in 
the formulae for $\delta D$, $\delta \lamp$, $\delta \lambp$, $\delta \lamm$ and $\delta \lambm$ extra $\pm i$ 
factors appear in the commutators. This factors also appear in the dimensional reduction of the formulae of \cite{W-B} and, we believe, 
should be there. Of course in the abelian case this makes no difference, so our formulas agree with \cite{W4}.

\end{document}